\documentclass[final,1p,times]{elsarticle}

\newcommand{\ba}{{\bf a}}

\newcommand{\bY}{{\,\partial Y}}
\newcommand{\haute}{\hat h}
\newcommand{\haut}{h}
\newcommand{\per}{\hat \ell}
\newcommand{\Per}{\ell}
\newcommand{\ff}{f}
\newcommand{\Ss}{\hat S}

\newcommand{\om}{{\Omega}}

\newcommand{\ee}{{\bm \varepsilon}}
\newcommand{\mI}{{\mathsf \bm I}}

\newcommand{\Ben}{D_\text{\tiny p}}

\newcommand{\Et}{E_\text{\tiny p}}
\newcommand{\Eg}{E_\text{\tiny s}}
\newcommand{\nut}{\nu_\text{\tiny p}}
\newcommand{\nug}{\nu_\text{\tiny s}}

\newcommand{\E}{\text{E}}
\newcommand{\C}{\text{C}}
\newcommand{\e}{\text{e}}
\newcommand{\cc}{\text{c}}

\newcommand{\kk}{\kappa}
\newcommand{\ch}{\text{ch}}
\newcommand{\sh}{\text{sh}}

\newcommand{\epun}{Z}

\newcommand{\kt}{\alpha_\text{\tiny T}}
\newcommand{\kkt}{k_\text{\tiny T}}

\newcommand{\ray}{\hat r}
\newcommand{\Yt}{Y}

\newcommand{\Zt}{Z}

\newcommand{\rp}{r_\text{\tiny p}}

\newcommand{\htt}{\ell_\text{\tiny p}}
\newcommand{\rt}{\rho_\text{\tiny p}}
\newcommand{\rg}{\rho_\text{\tiny s}}
\newcommand{\mut}{\mu_\text{\tiny p}}
\newcommand{\mug}{\mu_\text{\tiny s}}
\newcommand{\lamt}{\lambda_\text{\tiny p}}
\newcommand{\lamg}{\lambda_\text{\tiny s}}

\newcommand{\bu}{{\bf u}}
\newcommand{\bv}{{\bf v}}
\newcommand{\sig}{{\bm \sigma}}
\newcommand{\btau}{{\bm \tau}}

\newcommand{\bx}{{\bf x}}

\newcommand{\bz}{{\bf z}}

\newcommand{\grad}{\nabla}
\renewcommand{\div}{\text{div}}

\newcommand{\dd}{\text{d}}

\newcommand{\ep}{\eta}

\newcommand{\toutin}{\left\{\begin{array}{l}}
\newcommand{\toutind}{\left\{\begin{array}{ll}}
\newcommand{\toutint}{\left\{\begin{array}{lll}}
\newcommand{\toutout}{\end{array}\right.}

\newcommand{\eu}{{\bm e}_1}
\newcommand{\ed}{{\bm e}_2}

\newcommand{\eal}{{\bm e}_\alpha}

\newcommand{\drs}[2]{\partial_{#2}{ #1}}

\newcommand{\dr}[2]{\frac{\partial #1}{\partial #2}}
\newcommand{\drd}[2]{\frac{\partial^2 #1}{\partial #2 ^2}}
\newcommand{\drt}[2]{\frac{\partial^3 #1}{\partial #2^3}}
\newcommand{\drq}[2]{\frac{\partial^4 #1}{\partial #2^4}}

\newcommand{\dz}{{\text d}z}
\newcommand{\cphi}{\varphi}

\newcommand{\rll}{R_\text{\tiny LL}}
\newcommand{\rlt}{R_\text{\tiny LT}}
\newcommand{\rtl}{R_\text{\tiny TL}}
\newcommand{\rtt}{R_\text{\tiny TT}}

\newcommand{\phii}{\phi^\text{\tiny inc}}
\newcommand{\psii}{\psi^\text{\tiny inc}}
\newcommand{\be}{\beta}

\newcommand{\aph}{A_\text{\tiny L}^\text{\tiny inc}}
\newcommand{\aps}{A_\text{\tiny T}^\text{\tiny inc}}

\newcommand{\kl}{\alpha_\text{\tiny L}}
\newcommand{\kkl}{k_\text{\tiny L}}

\newcommand{\tl}{\theta_\text{\tiny L}}
\renewcommand{\tt}{\theta_\text{\tiny T}}

\newcommand{\kn}{\xi}
\newcommand{\au}{a}

\newcommand{\bo}{{\bf 0}}
\newcommand{\bua}{{\bf w}}

\newcommand{\ua}{ w}
\newcommand{\bsa}{{\bm \pi}}
\newcommand{\sa}{\pi}
\newcommand{\Ua}{ W}

\newcommand{\mo}[1]{\overline{ #1 }}

\newcommand{\dsp}{\displaystyle}
\def\beq{\begin{equation}}
\def\eeq{\end{equation}}

\usepackage{lineno,hyperref}
\usepackage{amsmath,bm}
\journal{Journal of the Mechanics and Physics of Solids }

\begin{document}

\begin{frontmatter}

\title{Effective model for elastic  waves propagating in a substrate supporting a
dense  array of plates/beams with flexural resonances}

\author{Jean-Jacques Marigo}
\address{Laboratoire de M\'ecanique des solides, Ecole Polytechnique, \\Route de Saclay, 91120 Palaiseau, France}
\author{Kim Pham}
\address{IMSIA, ENSTA ParisTech - CNRS - EDF - CEA, Universit\'e Paris-Saclay, \\
828 Bd des Mar\'echaux, 91732 Palaiseau, France
}

\author{Agn\`es Maurel}
\address{Institut Langevin, ESPCI  ParisTech, CNRS,\\
1 rue Jussieu, Paris 75005, France}

\author{S\'ebastien Guenneau}
\address{Aix Marseille Univ, CNRS, Centrale Marseille, Institut Fresnel, Marseille, France }

\begin{abstract}
 We consider the effect of an array of plates or beams over  a semi-infinite elastic ground on the propagation of elastic waves hitting the interface. The plates/beams are slender bodies with  fle\-xu\-ral  resonances at low frequencies  able to perturb significantly the propagation of waves in the ground. An effective model is obtained using   asymptotic analysis and homo\-ge\-ni\-zation techniques, which  
 can be expressed in terms of the ground alone with  effective dynamic  (frequency-dependent) boundary conditions of the Robin's type. For an incident plane wave at oblique incidence, the displacement fields and  the reflection coefficients are obtained in closed forms and their validity  is inspected by comparison with direct numerics
  in a two-dimensional setting. 
\end{abstract}


\end{frontmatter}


\section{Introduction}

We are interested in  wave propagation in a semi-infinite elastic substrate supporting a perio\-dic and dense array of thin or slender bodies. This is the  canonic idealized   configuration  used  
to illustrate the problem of "site-city interaction". Such a problem, recent on the seis\-mo\-lo\-gy history scale,   aims to 
 account for the urban environment as a factor modifying the seismic ground motion.
Starting  in the 19$^\text{th}$ century, the interest was  primarily 
 focused on the motion of the soil elicited by static or dynamic sources being concentrated 
 or distributed  on the free surface in the absence of buildings. 
These studies  have led to   important results as the Lamb's problem \cite{lamb,pekeris}. 
Then, more realistic configurations have been considered using approximate  models 
  to predict the effect of complex soils, 
including the presence of buried foundations, on the displacements in structures on the ground, see {\em e.g.}
 \cite{jennings1973,kausel78,gazetas1991} and \cite{kausel2010} for a review. 
In the classical  two-step model,  the displacements in the soil without structures above, so-called free fields, were firstly calculated 
  and   they were subsequently used as input data to determine 
 the motion within the  structure \cite{kausel78,kausel2010}.  
This means that the interaction, refereed to as the soil-structure interaction, was restricted to the effect of
the soil on the  structure. 
In the mid-1970s, Luco and Contesse \cite{luco1973}  and Wong and Trifunac  \cite{wong1975} studied the interaction between nearby buildings and they evidenced the resulting  modification on the ground motion. They termed this mutual interaction the structure-soil-structure interaction, which has been later renamed  soil-structure-soil interaction. On the basis of  these pioneering works the idea took root that several structures may interact with each other and  modify the ground motion, supplied by numerical simulations and direct observations during earthquakes   
\cite{wirginBard96,gueguen2000,clouteau2001,tsogka2003,gueguen2005soil,semblat2006,groby2008,uenishi2010}.   
At the scale of a city with the specificity of the presence of a sedimentary basin, the soil-structure-soil interaction   has been called site-city interaction, a term firstly coined bu Gu\'eguen \cite{gueguen2000}. 
From a theoretical point of view, most of the models encapsulate the response of a building 
with a  single or multi-degree of freedom system  
\cite{todorovska1992,gueguen2000,gueguen2002,boutin2004,ghergu2009,uenishi2010}. On the basis of this model,  
Boutin, Roussillon and co-workers have developed homogenized models  where the multiple interactions between periodically located oscillators are accounted for  from a macroscopic city-scale point of view
 \cite{boutin2004,boutin2006,boutin2013,boutin2015,boutin2016}. In the low frequency limit,  that is when the incident wavelength is large compared to the resonator 
spacing, the effect of the resonators can be encapsulated in
effective boundary conditions of the Robin type  for the soil, a result that we shall recovered in the present study.  
Such a mass-spring  model  
has been  used in physics  for randomly distributed oscillators \cite{garova1999} and periodically distributed oscillators \cite{wegert2010,vitali2015,maznev2018} for their influence  on 
 surface Love and Rayleigh waves.
The ability of arrays of resonators to block Love and Rayleigh waves has been exploited to envision new devices of seismic metasurfaces   \cite{kuznetsov2011,meta2012,meta2013,meta2014,krodel2015,meta2016,colquitt,nousLove,palermo2018a,palermo2018b} in analogy with  metasurfaces in acoustics \cite{kelders1998,rainbow2013} and in electromagnetism
 \cite{pendry2004,garcia2005}.

 \begin{figure}[t!]
\centering
\includegraphics[width=1\columnwidth]{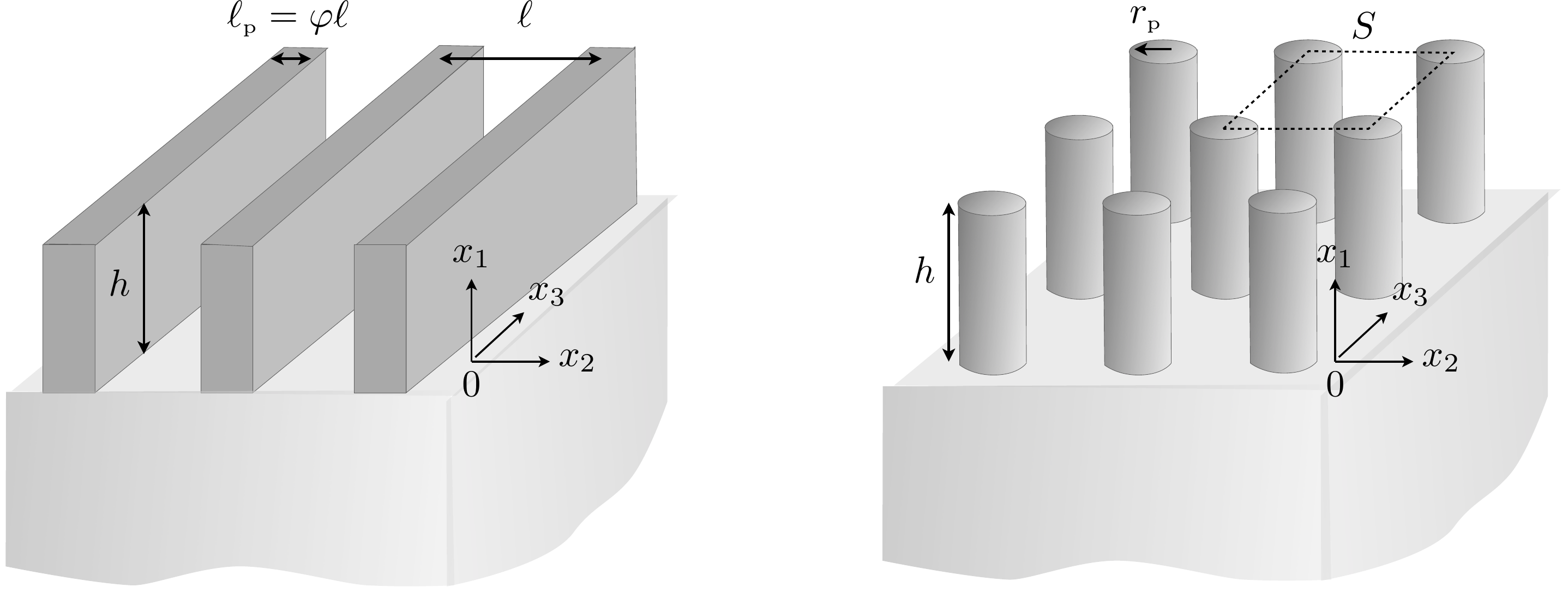}
\small (a) \hspace{7cm} (b)
	\caption{ Geometry of the actual problems: 
	(a) Array of parallel plates infinite along $x_3$ atop an isotropic substrate; 
	(b) Doubly periodic array of cylindrical beams atop an isotropic substrate.
The scalings are chosen to capture  the bending resonances only; with $\kkt$ the  wavenumber in the substrate, $\kkt\haut=\ep$ and $\kkt\Per=O(\ep^2)$.  }
\label{Fig1}
\end{figure}

In this study, we use asymptotic analysis and homogenization techniques to revisit the problem of the interaction of  a periodic array of plates or beams on the propagation of seismic waves in three dimensions. We consider slender bodies in the low frequency limit which means two things. Firstly, the typical wavelength $1/k$ is much larger than the array spacing $\ell$, which is a classical hypothesis. Secondly, we focus on the lowest resonances of the bodies being flexural resonances.  
 The first flexural resonances correspond to
  $kh\sim \htt/h$, with   $\htt$ the body thickness, $h$  the body height and 
   $\htt/h$ the slenderness (in comparison the first longitudinal resonance appears at $kh\sim 1$).
Now, we consider dense arrays, which means that  $\htt\sim\Per$, and $\cphi=\htt/\Per\in(0,1)$ (Figure \ref{Fig1}). 
Hence the asymptotic analysis is conducted considering that 
$$ \text{the wavelength $1/k$ is large compared to  $h$ which is itself large compared to  $\htt\sim \Per$.}
 $$
It is worth noting that assuming $\cphi=O(\ep^n)$ with $n\geq 1$ would allow  a reduction of model in a first step, resulting in  concentrated force problems, as implicitly considered in \cite{boutin2004,boutin2006,colquitt}. 
Here on the contrary, the  implementation of the asymptotic method will require that we reconstruct the asymptotic theory of plates and beams in a  low frequency regime, as previously done for a single body in solid mechanics, see {\em e.g.} \cite{ciarlet79,caillerie82} for plates and  \cite{trabucho87,geymonat87,miara90,corre2018} for beams. However, this classical theory has to be complemented with matched asymptotic expansions to link the behavior in the periodic set of  bodies with that in the substrate. 
This ``soil-structure" coupling requires a specific treatment   as used in interface homogenization \cite{nous1,nous2,nous3,nous4}, see also \cite{noushelmholtz} for a resonant case. 
In the present case, we shall derive effective boundary and transmission conditions applying in a homogenized region which replaces the actual array; and in this effective region the wave equation for flexural waves applies. This problem can be further simplified in  effective boundary conditions  of the Robin type applying on the surface of the soil, namely 
\beq
\sig\cdot {\bf n}={\mathsf K}(\omega)\bu,
\eeq 
where the frequency-dependent rigidity matrix ${\mathsf K}$  depends explicitly on the flexural frequencies of the plates/beams.  The rigidity matrix is diagonal as soon as the bodies have sufficient symmetry, resulting in effective impedance conditions which ressemble  those
obtained in \cite{boutin2013} in the same settings. 

%

\vspace{.3cm}
The paper is organized as follows. In \S \ref{sec2}, we summarize the   result of the asymptotic analysis in the case of an array of plates, whose detailed derivation is given in the \S \ref{sec3}. The resulting "complete" formulation \eqref{CL}-\eqref{CT1}   is equivalent to that in \eqref{CT2}-\eqref{deff} thanks to a partial  resolution of the problem. In \S \ref{secnum}, the accuracy of the effective model is inspected  by comparison with direct numerics based on multimodal method \cite{petitmonstre} for an in-plane incident wave. The strong coupling of the array with the ground at the flexural resonances are exemplified and the agreement between the actual and effective problems is discussed. We finish the study
 in \S \ref{conclude}  with concluding remarks and perspectives. 
We provide in the appendix \ref{appM}  the effective problem for the  an array of beams which is merely identical to the case of the plates with some specificities which are addressed.


\section{The actual problem and the effective problem}
\label{sec2}

We  consider in this section the asymptotic analysis of  an array of parallel plates atop an isotropic elastic substrate. 
We note that the problem splits in the in-plane and out-of-plane polarizations. 
The latter case has been already addressed in \cite{nousLove}. 
We focus in this section on the former, in-plane, case. 
We further note that the asymptotic analysis of  a doubly periodic array of cylinders atop an isotropic substrate is a fully coupled elastodynamic wave problem, which is thus slightly more involved and addressed in the Appendix.  

\subsection{The physical problem}
We consider the equation of elastodynamics for the displacement vector $\bu$ 
 the stress tensor $\sig$ and the strain tensor $\ee$
\beq\label{phys}\toutind
\text{in the  substrate}, x_1\in(-\infty,0): &  \div \sig+\rg \omega^2 \bu=\bo,
\quad \sig=2\mug \ee+\lamg \text{tr}(\ee) \mI, \quad \ee=\frac{1}{2} (\grad \bu+\;^t\grad \bu), \\[10pt]
\text{in the plates}, x_1\in(0,\ell): &  \div \sig+\rt \omega^2 \bu=\bo,\quad \sig=2\mut \ee+\lamt \text{tr}(\ee) \mI, 
\toutout
\eeq
 with  the Lam\'e coefficients $(\lamt,\mut)$ of the plates and $(\lamg,\mug)$ of the substrate, 
 $\omega$ the angular frequency and $\mI$ stands for the identity matrix. 
In three dimensions with $\bx=(x_1,x_2,x_3)$, stress free conditions $\sig\cdot {\bf n}=\bo$ apply at each boundary between an elastic medium (the plates  or the substrate) and  air, with $\bf n$ the normal to the interface. Eventually, the continuity of the displacement and  the normal stress apply at each boundary between the parallel  plates and the substrate.
This problem can be solved once the source $\bu^\text{\tiny inc}$ has been defined and accounting for the radiation condition for $x_1\to-\infty$ which applies to the scattered field $(\bu-\bu^\text{\tiny inc})$.

\subsection{The effective problem}

\begin{figure}[h!]
\centering
\includegraphics[width=1\columnwidth]{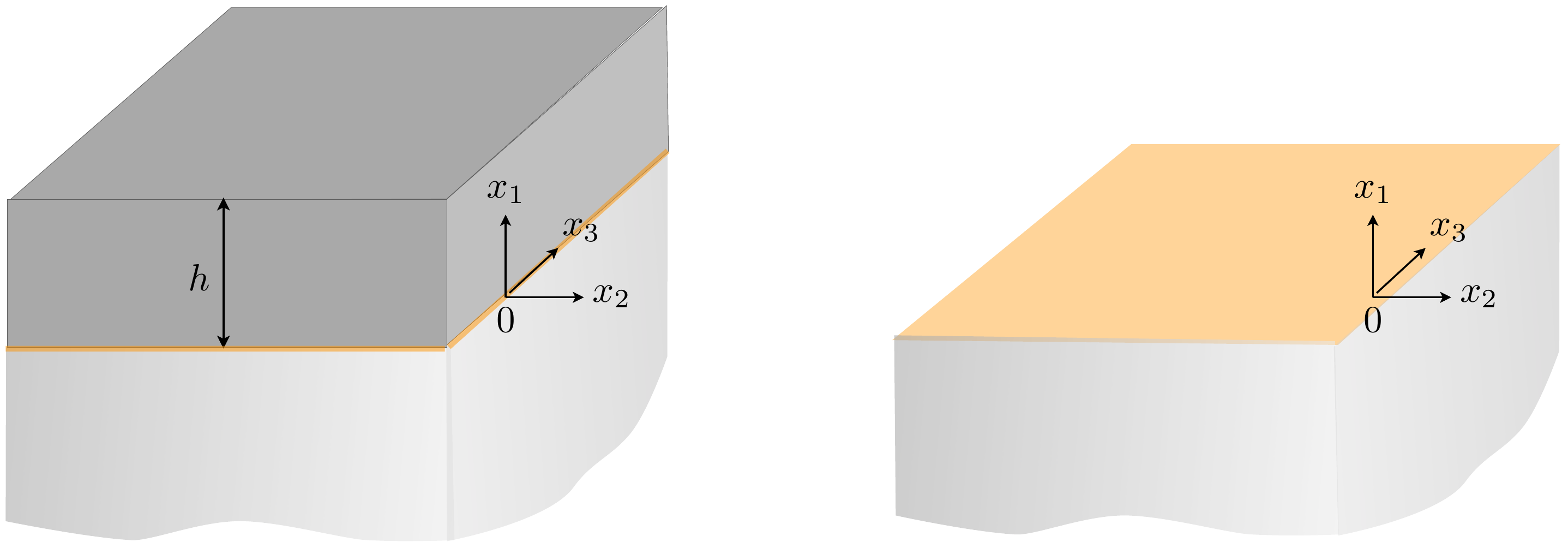}

\hspace{.1cm}
\small (a) \hspace{7cm} (b)
	\caption{ Geometry in the effective problems for an array of plates. (a) In the complete formulation, the region of the array is replaced by an equivalent layer where  \eqref{CL} applies, complemented by the transmission conditions in \eqref{CT1}. (b) In the impedance formulation, the problem is reduced to effective boundary conditions \eqref{CT2}  which hold at $x_1=0$. The same holds for the array of beams, with (a)  \eqref{complet3}-\eqref{CT3d}  and (b)  \eqref{CT3p}. }
	\label{Fig02}
\end{figure}

Below we summarize the main results of the analysis developed in the  \S\ref{sec3} and which provides the so-called "complete formulation" where the array of parallel plates is replaced by an e\-qui\-va\-lent layer associated to effective boundary and transmission conditions (Figure \ref{Fig02}(a)). Owing to a partial resolution, this formulation can be simplified to an equivalent "impedance formulation" set on the substrate only (Figure \ref{Fig02}(b)).
 We note that all three components of the displacement field appear in this section, and the reader should be aware that we make use of variables $\bx=(x_1,x_2,x_3)$ and $\bx'=(x_2,x_3)$.

\subsubsection{Complete formulation}
The effective problem reads as follow
\beq\label{CL}\toutind
\dsp \text{In the substrate}, x_1\in(-\infty,0): & \div \sig+\rg \omega^2 \bu=\bo,  \quad \sig=2\mug \ee+\lamg \text{tr}(\ee) \mI,  \\[12pt]
\dsp \text{In the region of the plates}, x_1\in(0,\haut): &\dsp  \drq{u_2}{x_1} -\kk^4\, u_2=0,\quad \kk=\left(\frac{\rt \omega^2\htt}{\Ben}\right)^{1/4},\\[10pt]
& u_{1}(x_1,\bx')=u_{1}(0,\bx'),\quad u_3(x_1,\bx')=u_3(0,\bx'), 
\toutout
\eeq
 with $\bx'=(x_2,x_3)$, 
\beq
\Ben=\frac{\Et}{(1-\nut^2)}\frac{ \htt^3}{12},\eeq
 the flexural rigidity of the plates ($\rt$ the mass density, $\Et$ the Young's modulus and $\nut$ the Poisson's ratio), complemented with   boundary conditions at  $x_1=0$ and $x_1=\haut$ of the form
\beq\label{CT1}
\toutin
\dsp \sigma_{11}(0^-,\bx')=\rt\omega^2\cphi \haut \; u_1(0,\bx'),\quad
\dsp \sigma_{12}(0^-,\bx')=-\frac{\Ben}{\Per}\,  \;\drt{u_2}{x_1}(0^+,\bx'), 
\\[10pt]
\dsp \sigma_{13}(0^-,\bx')=\cphi\haut\left(\Et \drd{u_3}{x_3}(0^-,\bx')+\rt \omega^2 u_3(0^-,\bx')\right), 
\\[10pt]
\dsp {u_2}(0^+,\bx')={u_2}(0^-,\bx'),\quad \dsp \dr{u_2}{x_1}(0^+,\bx')=0,
\\[10pt]
\dsp \drd{u_2}{x_1}(\haut,\bx')=\drt{u_2}{x_1}(\haut,\bx')=0.
\toutout\eeq
These effective conditions express (i) at $x_1=0$ a balance of the stress, prescribed displacements and vanishing  rotation and (ii) at $x_1=\haut$, free end conditions with vanishing  bending moment and shearing force.
One notes that all three components of the displacement field $\bu$ appear in \eqref{CT1} which involves partial derivatives on $x_1$ and $x_3$ only.

\subsubsection{Impedance formulation}

From \eqref{CL}, the problem  in $x_1\in (0,\haut)$ can be solved owing to the linearity of the problem  with respect to $u_2(0^-,x_{2})$, see  \ref{aparbre}.
Doing so, the  problem can be thought in the substrate only along with the boundary conditions of the Robin's type, namely
\beq \label{CT2}
\toutin
\dsp \div \sig+\rg \omega^2 \bu=\bo,  \quad \sig=2\mug \ee+\lamg \text{tr}(\ee) \mI,  \quad x_1\in(-\infty,0),
\\[12pt]
\dsp \sigma_{11}(0,\bx')=\; \epun \,u_1(0,\bx'),	  \quad
\dsp \sigma_{12}(0,\bx')
=\epun\;f(\kk\haut) \; u_2(0,\bx'),\\[10pt]
\dsp \sigma_{13}(0,\bx')=\cphi\haut\left(\Et \drd{u_3}{x_3}(0,\bx')+\rt \omega^2 u_3(0,\bx')\right),
\toutout\eeq  
with the following impedance parameters
\beq\label{deff}
\epun=\rt\omega^2\cphi \haut,\quad f(\kk \haut)=\frac{\sh\kk\haut \cos\kk\haut+\ch\kk\haut \sin\kk\haut}{\kk\haut(1+\ch\kk\haut\cos\kk\haut)},
\eeq
(we have used that $\Ben \kk^4=\rt \omega^2\htt$). The   conditions on $(\sigma_{11},\sigma_{12})$   encapsulate  
 the effects of the in-plane bending  of the plates while the condition on $\sigma_{13}$
 can be understood as the equilibrium of an axially loaded bar  (in the absence of substrate, we recover the wave equation for quasi-longitudinal waves).
  It is worth noting that for out-of-plane displacements, $u_3(x_1,x_2)$ and $u_1=u_2=0$, the  boundary conditions simplify to  $\sigma_{13}(0,x_2)=\rt \cphi\haut \omega^2 u_3(0,x_2)$. This corresponds to the impedance condition which can be deduced
   from the analysis conducted in \cite{nousLove} and resulting in $\sigma_{13}(0,x_2)=\mut\cphi \kkt\tan(\kkt\haut) u_3(0,x_2)$ and obtained here in the limit $\kkt\haut \ll 1$.
  
  \section{Derivation of the effective problem}
  \label{sec3}
  As previously said, the asymptotic analysis is conducted considering that 
the typical wavelength $1/k$ is large compared to  the plate height $h$ which is itself large compared to the array spacing $ \Per\sim \htt$.
Hence, with  $\kkt=\omega\sqrt{\rg/\mug}$ and  $\kkl=\omega\sqrt{\rg/(\lamg+2\mug)}$ of the same order of magnitude,
 we define the small non-dimensional  parameter $\ep$ as
\beq\nonumber
\ep=\sqrt{\kkt \Per},\quad \text{and} \quad \kkt\haut=O(\ep),
\eeq
(note that to excite both the bending and the longitudinal modes another scaling is required with  $k\haut=O(1)$, and this is a higher frequency regime studied in  \cite{colquitt}).
 Accordingly, the asymptotic analysis is conducted  using the rescaled height $\haute$ of the plates and array's spacing $\per$ defined by
\beq\nonumber
(\haute,\per)=\left (\frac{\haut}{\ep},\frac{\Per}{\ep^2}\right),
\eeq
which models an array of densely packed thin plates. We also define the associated rescaled spatial coordinates
\beq\label{coor}
y_1=\frac{x_1}{\ep}, \quad \bz=\frac{\bx}{\ep^2}. 
\eeq

\subsection{Effective wave equation in the region of the plates } 
 
 \subsubsection{Notations}
In the region of the array of parallel plates, the displacements and the stresses vary in the horizontal direction over small distances dictated by $\Per$, and over large distances  dictated by the incoming waves; these two scales are accounted for by the  two-scale coordinates $(\bx',z_2)$, with $\bx'=(x_2,x_3)$. In the vertical direction, the variations are dictated by $\haut$ only and this is accounted for by the rescaled coordinate $y_1$. It follows that the fields $(\bu,\sig)$ are written of the form
\beq\label{exp}
\bu=\sum_{n\geq0}\ep^n\bua^n(y_1,z_2,\bx'), \quad \sig=\sum_{n\geq0}\ep^n\bsa^n(y_1,z_2,\bx'),
\eeq
with 
the three-scale differential operator  reading
\beq\label{diff}
\grad\to \frac{\eu}{\ep}\dr{}{y_1}+ \frac{\ed}{\ep^2}\,\dr{}{z_2}+\grad_{\bx'},
\eeq
where $\eu=(1,0,0)$ and $\ed=(0,1,0)$.
Now,  we  introduce the   strain tensor   with respect to $\bx'$
\beq
\ee^{\bx'}(\bu)=\frac{1}{2}\begin{pmatrix} 
0 & \dsp\drs{u_1}{x_2} & \drs{u_1}{x_3} \\[8pt]
\drs{u_1}{x_2}  & 2\drs{u_2}{x_2} & \left(\drs{u_2}{x_3}+\drs{u_3}{x_2}\right) \\[8pt]
\drs{u_1}{x_3} & \left(\drs{u_2}{x_3}+\drs{u_3}{x_2}\right) & 2\drs{u_3}{x_3}  
\end{pmatrix},
 \eeq
and the strain tensors with respect to the rescaled coordinates $y_1$ and $z_2$,
 \beq
\ee^{y_1}(\bu)=\frac{1}{2}\begin{pmatrix} 
2\drs{w_1}{y_1} & \drs{u_2}{y_1} & \drs{u_3}{y_1} \\[8pt]
 \drs{u_2}{y_1}  & 0 & 0 \\[8pt]
 \drs{u_3}{y_1}  & 0 & 0 
\end{pmatrix},\quad
 \ee^{z_2}(\bu)=\frac{1}{2}\begin{pmatrix} 
0 & \drs{u_1}{z_2} & 0 \\[8pt]
\drs{u_1}{z_2} & 2\drs{u_2}{z_2} & \drs{u_3}{z_2} \\[8pt]
0 & \drs{u_3}{z_2} & 0  
\end{pmatrix}.
 \eeq

 The system in the region of the plates reads, from \eqref{phys},
 \beq\label{eq13d}\toutind
(\E_{1})\quad \dsp \frac{1}{\ep}\drs{\sigma_{11}}{y_1}+\drs{\sigma_{1\alpha}}{x_\alpha}+\frac{1}{\ep^2}\,\drs{\sigma_{12}}{z_2}+\rt \omega^2u_1=0,\\[10pt]
(\E_{\alpha})\quad \dsp \frac{1}{\ep}\drs{\sigma_{\alpha 1}}{y_1}+\drs{\sigma_{\alpha\beta}}{x_\beta}+\frac{1}{\ep^2}\,\drs{\sigma_{\alpha 2}}{z_2}+\rt \omega^2u_\alpha=0,\\[10pt]
 (\C)\quad \dsp \sig=\frac{1}{\ep}\left(2\mut \ee^{y_1}+\lamt \text{tr}(\ee^{y_1})\right)+\left(2\mut \ee^{\bx'}+\lamt \text{tr}(\ee^{\bx'})\right)+\frac{1}{\ep^2}\left(2\mut \ee^{\bz'}+\lamt \text{tr}(\ee^{\bz'})\right), \\[10pt]
 \toutout\eeq
with the convention on the Greek letters $\alpha=2,3$, the same for $\beta$, and where $\ee$ stands for $\ee(\bu)$.
 \begin{figure}[b!]
\centering
\includegraphics[width=.3\columnwidth]{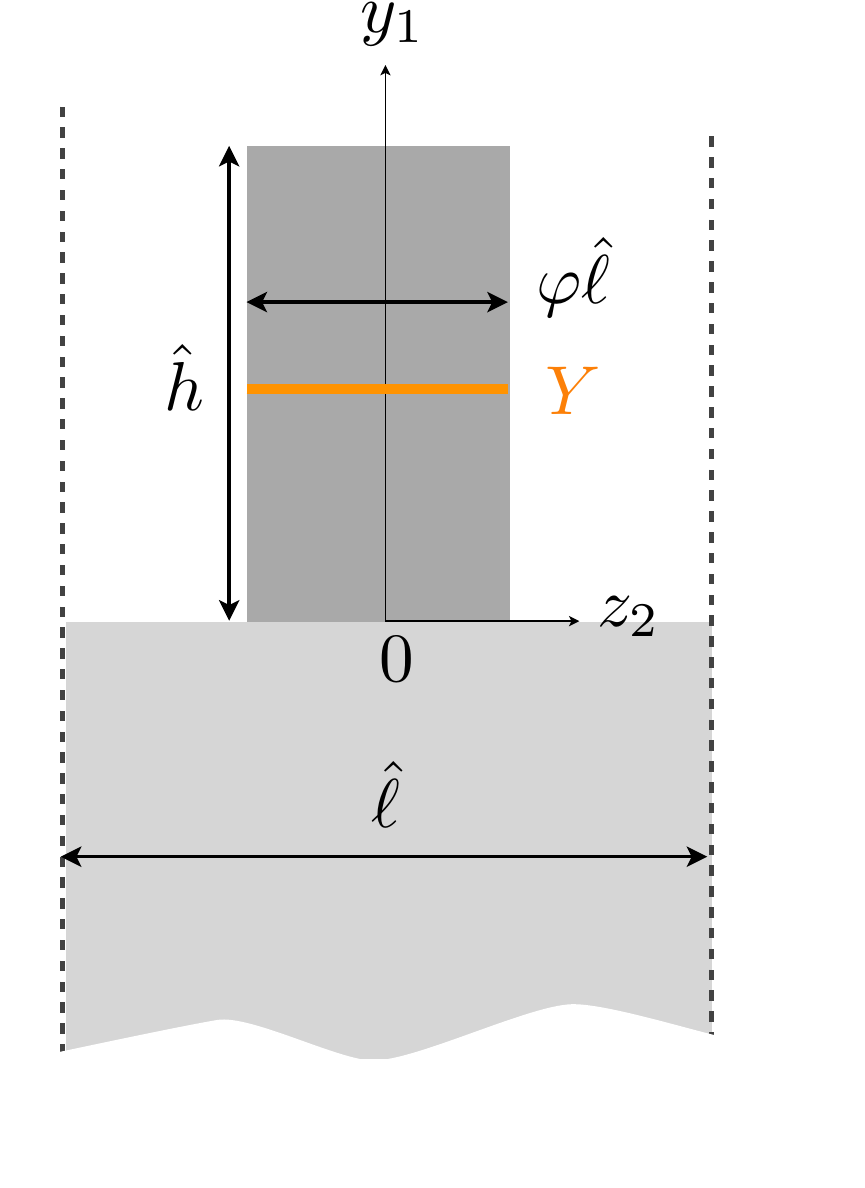}
	\caption{Analysis of a single plate in  rescale coordinates,   \eqref{exp} with $Y=\{z_2\in(-\cphi\per/2,\cphi\per/2)\}$;
	 the analysis holds within the plate far from its boundaries at $y_1=0,\haute$.}
	\label{Fig2}
\end{figure}
We shall use  the stress-strain relation written in the form
\beq\label{trez}
 (\C')\quad \dsp \frac{1}{\ep}\ee^{y_1}+\ee^{\bx'}+\frac{1}{\ep^2} \,\ee^{z_2}=\frac{(1+\nut)}{\Et}\, \sig-\frac{\nut}{\Et}\,\text{tr}(\sig)\,\mI.
 \eeq
 Eventually, the boundary conditions read
 \beq\label{dt}
 {\sigma_{2i}}=0, \quad i=1,2,3, \quad {at } \; z_2=\pm\cphi \per/2,
 \eeq
and are complemented by boundary conditions at $y_1=0,\haute$ assumed to be known (they will be justified later).
We seek to establish the effective behaviour in the region of the array in terms of macroscopic averaged fields which do not depend anymore on the rapid coordinate $z_2$ associated with the small scale  $\per$ as the following averages taken along rescaled variable $z_2$. These fields are defined at any order $n$ as 
\beq\nonumber
\mo{\bua^n}(y_1,\bx')=\frac{1}{\cphi\per}\int_Y\bua^n(y_1,z_2,\bx')\,\dz_2, \quad \mo{\bsa^n}(y_1,\bx')=\frac{1}{\per}\int_Y\bsa^n(y_1,z_2,\bx')\,\dz_2,
\eeq
with  $Y=\{z_2\in(-\cphi\per/2,\cphi\per/2)\}$ the segment shown in figure \ref{Fig2}.

\subsubsection{Sequence of resolution and main results of the analysis}
We shall derive the equation satisfied in the region of the array, and additional results on the stresses $(\mo{\sa^0_{1i}},\mo{\sa^1_{1i}}\,)$, $i=1,2,3$,  required to establish the effective boundary conditions at $y_1=0,\haute$. The main results will be  obtained following the procedure : 

\begin{enumerate}
\item{ We establish the following properties on $\bsa^0$ 
\beq\label{kim1}
\mo{\sa^0_{11}}=0, \quad \sa^0_{2i}=0,\; i=1,2,3, \quad \mo{\sa_{13}^0}=0.
\eeq 
}
\item{ Then we derive the dependence of $(\bua^0,\bua^1)$ on $z_2$ which have the form
\beq\label{kim2}\toutin
\ua^0_1=W^0_1(\bx'), \quad \ua^0_2=W^0_2(y_1,\bx'),\quad \ua^0_3=W^0_3(\bx'),\\[10pt]
\dsp \ua^1_1=W_1^1(\bx')-\frac{\lamt}{2(\lamt+\mut)}\dr{W_3^0}{x_3}(\bx')\,y_1 - \dr{W^0_2}{y_1}(y_1,\bx')\,z_2, \quad \ua^1_\alpha=W^1_\alpha(y_1,\bx'),\\
\toutout\eeq
and 
\beq\label{kim3}
\dsp  \sa_{11}^0=-\frac{\Et}{1-\nut^2}\drd{W^0_2}{y_1}(y_1,\bx')\; z_2,\quad
\dsp  \mo{\sa_{33}^0}=\cphi \Et\dr{W_3^0}{x_3}(\bx').
 \eeq
}
\item{ Eventually, we identify the form of $\mo{\sa^1_{1i}}$, $i=1,2,3$, 
and the Euler -Bernoulli equation governing the bending $W^0_2$. Specifically
\beq\label{kim4}\toutin
\dsp \mo{\sa_{11}^1}(y_1,\bx')=\rt \omega^2\cphi\, W_1^0(\bx')\,(\haute-y_1),\\[12pt]
\dsp \mo{\sa^1_{12}}(y_1,\bx')=-\frac{\Et}{(1-\nut^2)}\frac{\cphi^3\per^2}{12}\drt{W_2^0}{y_1}(y_1,\bx'),\\[12pt]
\dsp \mo{\sa^1_{13}}(y_1,\bx')=\cphi \left(\Et\drd{W^0_3}{x_3}(\bx')+\rt\omega^2 W^0_3(\bx')\right)\left(\haute-y_1\right),
\toutout\eeq
and 
\beq\label{kim5}
\frac{\Et}{(1-\nut^2)}\,\frac{\cphi^2\per^2}{12}\, \drq{\Ua^0_2}{y_1} -\rt \omega^2\,\Ua^0_2=0.
\eeq
}
\end{enumerate}
In the remainder of this section, we shall establish the above results. 
We shall denote  $(\E_{1})^{n}$, $(\E_{\alpha})^{n}$ and $(\C)^{n}$ the equations 
which correspond to terms in \eqref{eq13d}  factor of $\ep^n$.

\subsubsection{First step: properties of $\bf \pi^0$ in \eqref{kim1}}
From $(\E)^{-2}$ in \eqref{eq13d}, we have that $\drs{\sa^0_{2i}}{z_2}=0$, which along with the boundary conditions at $z_2=\pm\cphi\haute/2$ leave us with $\sa^0_{2i}=0$. Next from $(\E_1)^{-1}$ and $(\E_3)^{-1}$, we also have that $\drs{\sa^0_{11}}{y_1}+\drs{\sa^1_{12}}{z_2}=0$ and $\drs{\sa^0_{13}}{y_1}+\drs{\sa^1_{23}}{z_2}=0$; by averaging these relations over $Y$ and accounting for ${\sa^1_{2i}}_{|\bY}=0$, we get  that $\mo{\sa^0_{11}}$ and $\mo{\sa^0_{13}}$ do not depend on $y_1$. We now anticipate the boundary condition  $\mo{\sa^0_{11}}(\haute,\bx')=\mo{\sa^0_{13}}(\haute,\bx')=0$ that we shall prove later on (see forthcoming \eqref{gg}),  we get $\mo{\sa^0_{11}}=\mo{\sa^0_{13}}=0$ in $Y$. We have the properties announced in \eqref{kim1}.

\subsubsection{Second step:   $(\bua^0,  \bua^1)$ in \eqref{kim2} and $(\sa^0_{11},\mo{\sa_{33}^0}\,)$ in \eqref{kim3} }
Some of the announced results are trivially obtained.  From $(\C')^{-2}$ in \eqref{trez}, we get that $\drs{\ua^0_i}{z_2}=0$,  and from $(\C'_{11})^{-1}$ and $(\C'_{13})^{-1}$ that $\drs{\ua^0_1}{y_1}=\drs{\ua^0_3}{y_1}=0$, which  leaves us with the form of $\bua^0$ in \eqref{kim2}.
Next $(\C'_{\alpha\alpha})^{-1}$ tells us that 
 $\drs{\ua^1_\alpha}{z_2}=0$, in agreement with the form of $\ua_\alpha^1$ in \eqref{kim2}. We have yet to derive the form of $\ua_1^1$, which is more demanding. From $(\C'_{12})^{-1}$, $\drs{\ua^1_1}{z_2}=-\drs{\ua_2^0}{y_1}$ and thus
 \beq\label{inter1}
 \ua^1_1=W_1(y_1,\bx')-\dr{W^0_2}{y_1}(y_1,\bx')\, z_2,
 \eeq
 but we can say more on $W_1$. Let us consider the system provided by $(\C_{11})^0$ and $(\C_{22})^0$, specifically 
 \beq\toutind
 \sa^0_{11}=&\dsp (\lamt+2\mut)\drs{\ua_1^1}{y_1}+\lamt\left(\drs{\ua^0_\alpha}{x_\alpha}+\drs{w^2_2}{z_2}\right),\\
0=&\dsp (\lamt+2\mut)\left(\drs{\ua_2^2}{z_2}+\drs{\ua_2^0}{x_2}\right)+\lamt\left(\drs{w^1_1}{y_1}+\drs{\ua^0_3}{x_3}\right),
 \toutout\label{trytry}\eeq
where we have used that $\sa_{22}^0=0$. After elimination of $\drs{\ua_2^2}{z_2}$ and owing to the form of $\ua^0_\alpha$ in \eqref{kim1} and $\ua^1_1$ in \eqref{inter1} (at this stage), we get  $\sa^0_{11}=a(y_1,\bx')z_2+b(y_1,\bx')$ with  
\beq
a=-\frac{\Et}{1-\nut^2}\drd{W^0_2}{y_1}(y_1,\bx'),\quad b=\frac{2\mut}{(\lamt+2\mut)}\left(2(\lamt+\mut)\dr{W_1}{y_1}(y_1,\bx')+\lamt\dr{W_3^0}{x_3}(\bx')\right),
\eeq  
(we have used that ${\Et}/(1-\nut^2)=4\mut(\mut+\lamt)/(\lamt+2\mut)$). 
It is now sufficient to remark that $\mo{\sa_{11}^0}=0$ imposes $b=0$. This immediately provides the form of $\sa^0_{11}$ in \eqref{kim3} and  
 \beq
 W_1(y_1,\bx')=W_1^1(\bx')-\frac{\lamt}{2(\lamt+\mut)}\dr{W_3^0}{x_3}(\bx')\, y_1,
 \eeq
 which along with \eqref{inter1} leaves us with the form of $w_1^1$  in \eqref{kim2}.
The same procedure is used to get $\sa_{33}^0$, which from $(\C_{33})^0$, reads 
 \beq
 \sa^0_{33}=\dsp (\lamt+2\mut)\drs{\ua_3^0}{x_3}+\lamt\left(\drs{\ua^1_1}{y_1}+\drs{w^0_2}{x_2}+\drs{w^2_2}{z_2}\right).
\eeq
Using that $\sa_{22}^0=0$ to eliminate $\drs{w^2_2}{z_2}$, we get 
 \beq
 \sa^0_{33}=\dsp \Et \dr{W_3^0}{x_3}(\bx')-\frac{2\mut\lamt}{\lamt+2\mut}\drd{W_2^0}{y_1}(y_1,\bx')\,z_2,
\eeq
which after integration over $Y$ leaves us with $\mo{\sa^0_{33}}$ in \eqref{kim3}.
 Incidentally, $\ua^2_2$ can be determined from \eqref{trytry} and we find
 \beq\label{w22}
 \ua^2_2=-\left(\dr{W_2^0}{x_2}(y_1,\bx')+\frac{\lamt}{2(\lamt+\mut)}\,\dr{W^0_3}{x_3}(\bx')
 \right)z_2 +\frac{\lamt}{\lamt+2\mut}\, \drd{W_2^0}{y_1}\; \frac{z^2_2}{2}+W^2_2(y_1,\bx').
 \eeq

 \subsubsection{Third step: the $\mo{\sa^1_{1i}}$ in \eqref{kim4} and  the Euler-Bernoulli equation in \eqref{kim5}.}
We start with   $(\E)^0$ in \eqref{eq13d} integrated over $Y$, specifically,
\beq
\dsp \dr{\mo{\sa^1_{11}}}{y_1}+\rt\omega^2\cphi\,W_1^0=0,\quad
\dsp \dr{\mo{\sa^1_{12}}}{y_1}+\rt\omega^2\cphi\,W_2^0=0,\quad
\dsp \dr{\mo{\sa^1_{13}}}{y_1}+\dr{\mo{\sa^0_{33}}}{x_3}+\rt\omega^2W_3^0=0,
\label{yyy}\eeq
 where we have used \eqref{kim1} and $\bsa^2\cdot {\bf n}_{|\bY}=\bo$. Since $W^0_1$ and $W^0_3$ depend  on $\bx'$ only, and accounting for $\mo{\sa^0_{33}}(\bx')$ in \eqref{kim3}, we get by integration the forms of $\mo{\sa^1_{11}}$ and of $\mo{\sa^1_{13}}$ announced in \eqref{kim4}. Note that we have anticipated the boundary conditions   $\mo{\sa^1_{1i}}=0$ at $y_1=\haute$, see forthcoming \eqref{gg}.

 \vspace{.3cm}
 The equation on  $\mo{\sa^1_{12}}$ in \eqref{yyy} will provide the Euler-Bernoulli equation once $\mo{\sa^1_{12}}$ has been determined (the integration is not possible since $W^0_2$ depends on $y_1$). To do so, we use,  that $\drs{\sa^0_{11}}{y_1}+\drs{\sa^1_{12}}{z_2}=0$, from $(\E_1)^{-1}$,  along with $\sa^0_{11}$ in \eqref{kim3}. After integration and using the boundary condition of vanishing $\sa^1_{12}$ at $z_2=\pm\cphi\per/2$, we get that
\beq
 \sa^1_{12}=\frac{\Et}{2(1-\nut^2)}\drt{W_1^0}{y_1}(y_1,\bx')\,\left(z_2^2-\frac{\cphi^2\per^2}{4}\right),
\eeq
hence the form of  $\mo{\sa^1_{12}}$ in \eqref{kim4}. It is now sufficient to use  $\sa^1_{12}$ in \eqref{yyy} to get the Euler-Bernoulli announced in \eqref{kim5}.

\subsection{ Effective boundary conditions at the top of the array } 
\label{eff2}
\begin{figure}[h!]
\centering
\includegraphics[width=.7\columnwidth]{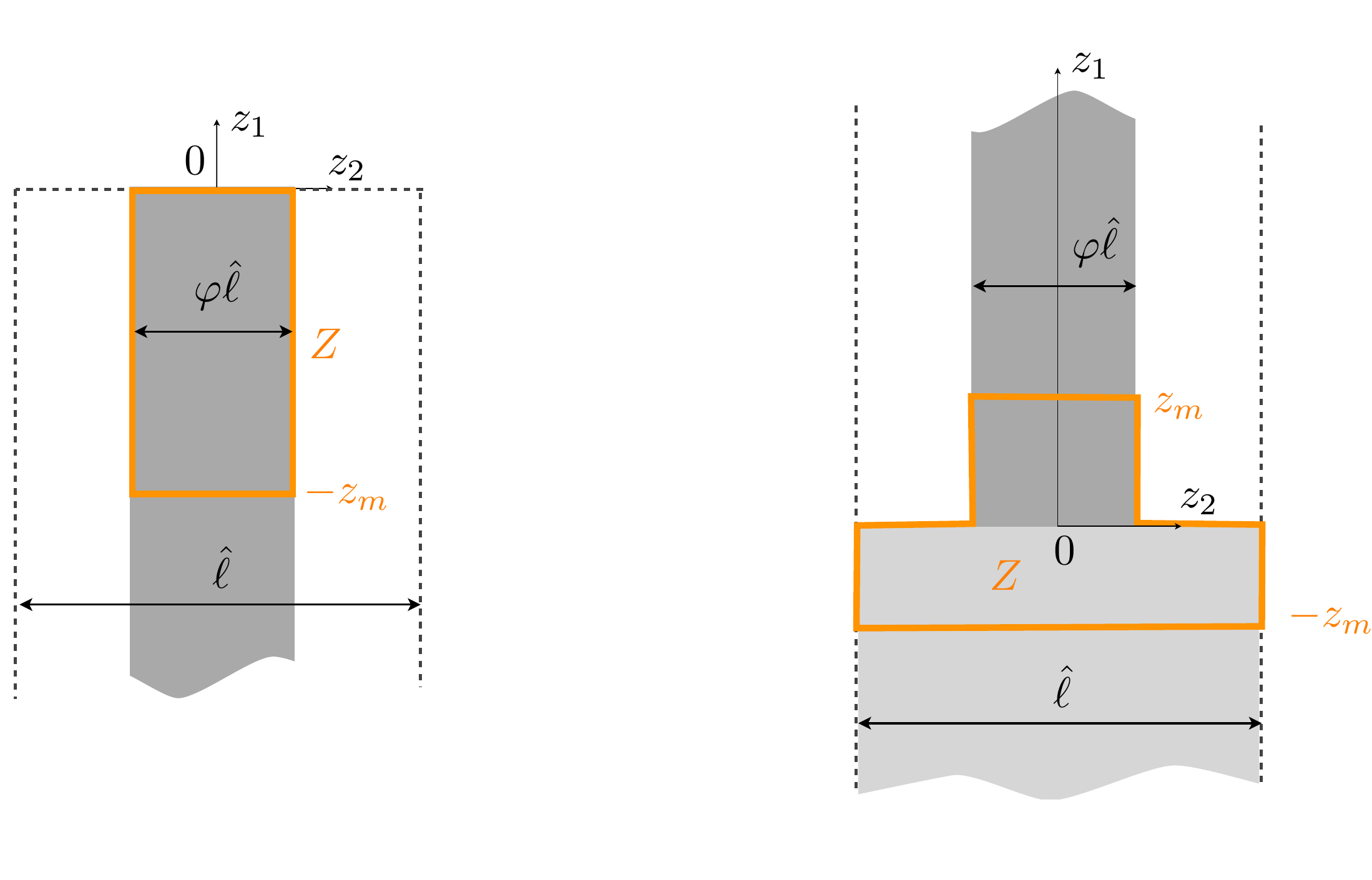}

\hspace{.1cm}\small (a) \hspace{5cm} (b)
	\caption{Analysis of the effective conditions at the top and at the bottom of the array. (a) Half strip at the top of the array in $(z_1,z_2)$ coordinate with $z_1\in (-\infty,0)$, (b)  Strip at the bottom of the array with $z_1\in (-\infty,+\infty)$; for $z_1\in(-\infty,0)$, the terms in the expansions \eqref{toto1}  are periodic with respect to $z_2\in (-\per/2,\per/2)$.}
	\label{Fig3}
\end{figure}

 To derive the transmission conditions at the top of the array, we perform a zoom  by substituting  $y_1$ used in \eqref{exp}  by $z_1=y_1/\ep$, see Figure \ref{Fig3}a. Accordingly, the expansions of the fields are sought of the form
 \beq\label{toto1}
\bu=\sum_{n\geq0}\ep^n\bv^n(\bz',\bx'), \quad \sig=\sum_{n\geq 0}\ep^n\btau^n(\bz',\bx'),
\eeq
where we denote $\bz'=(z_1,z_2)$.
The coordinate $z_1\in(-\infty,0)$ accounts for small scale variations  of the evanescent fields at  the top of the plates. Next, the boundary conditions will be obtained by matching the solution in \eqref{toto1}  for  $z_1\to - \infty$ with that in \eqref{exp} valid far from the boundary for  $y_1\to \haute$. This means that we ask the two expansions to satisfy 
\beq\nonumber
\bv^0(z_1,z_2,\bx')+\ep \bv^1(z_1,z_2,\bx')+\cdots \underset{z_1 \to -\infty}{\sim}  \bua^0(\haute+\ep z_1,z_2,\bx')+\ep \bua^1(\haute+\ep z_1,z_2,\bx')+\cdots ,
\eeq
(and the same for the stress tensors); note that we have used that $y_1=\ep z_1$.  It results that
\beq\label{dsd3}\toutind
\dsp \lim_{z_1\to -\infty}\bv^0(\bz',\bx')= \bua^0(\haute,\bx'),\quad &\dsp  \lim_{z_1\to -\infty}\left(\bv^1(\bz',\bx')-z_1\dr{\bua^0}{y_1}(\haute,z_2,\bx')\right)= \bua^1(\haute,z_2,\bx'),\\[10pt]
\dsp \lim_{z_1\to -\infty}\btau^0(\bz',\bx')= \bsa^0(\haute,z_2,\bx'), &\dsp \lim_{z_1\to -\infty}
\left(\btau^1(\bz',\bx')-z_1\dr{\bsa^0}{y_1}(\haute,z_2,\bx')\right)= \bsa^1(\haute,z_2,\bx').
\toutout
\eeq

According to the dependence of the fields in \eqref{toto1} on $(\bz',\bx')$, the differential operator  reads as follows
\beq\label{gradz}
\grad\to \frac{1}{\ep^2}\,\grad_{\bz'}+\grad_{\bx'},
\eeq
and we shall need only the first equation of \eqref{phys}, which reads 
\beq\label{eq11}
(\e)\quad \dsp \frac{1}{\ep^2}\,\div_{\bz'}\,\sig +\div_{\bx'}\sig+\rt \omega^2\,\bu=
{\bf 0},
\eeq
 where $\div_{\bz'}$ and $\div_{\bx'}$ means the divergence with respect to the coordinate $\bz'$ and $\bx'$ respectively.
 In \eqref{eq11}, $(\e)^{-2}$ and $(\e)^{-1}$ tell us that 
$\div_{\bz'} \btau^0=\div_{\bz'} \btau^1={\bm 0}$, that we
integrate over $Z=\{z_1\in(-z_m,0), z_2\in Y\}$ to get $\int_{\partial Z} \btau^0\cdot {\bf n} \;\dd s=\int_{\partial Z} \btau^1\cdot {\bf n}  \;\dd s=\bo$. On $\partial Z$, $\btau^0\cdot {\bf n}$ and $\btau^1\cdot {\bf n}$ vanish except on the bottom edge $Y$ of $\partial Z$ at $z_1=-z_m$ where ${\bf n}=-\eu$. It follows, from \eqref{dsd3} along with $\mo{\sa^0_{1i}}=0$ in \eqref{kim1}, that
\beq  \label{mmm}
\bo=\lim_{z_m\to \infty}\int_{\partial Z}\btau^0\cdot {\bf n}\,\dd s= -\per \,\mo{\sa^0_{1i}}(\haute,\bx')\,{\bm e}_i, \quad \bo=\lim_{z_m\to \infty}\int_{\partial Z}\btau^1\cdot {\bf n}\,\dd s= -\,\per \mo{\sa^1_{1i}}(\haute,\bx')\,{\bm e}_i.
\eeq
which provides the   boundary conditions
\beq\label{gg}
\mo{ \sa^0_{1i}}(\haute,\bx')=\mo{\sa^1_{1i}}(\haute,\bx')=0, \quad i=1,2,3.
\eeq
The  conditions on $\mo{\sa^0_{1i}}$ are consistent with \eqref{kim1}. 
The conditions on  $\mo{\sa^1_{11}}$ and $\mo{\sa^1_{13}}$ are those anticipated in the previous section, see \eqref{kim4}. 
Eventually, the condition $\mo{\sa^1_{12}}(\haute,x_2)=0$ combined with \eqref{kim4}  leads to the condition of zero shear force 
\beq
\drt{\Ua_2^0}{y_1}(\haute,\bx')=0.\label{jwx3}
\eeq
We have yet to derive the  condition of zero bending moment. First, integrating $\div_\bz \btau^0$ over $Z$, we get $0=\int_{\partial Z}\,\tau^0_{ij} n_{j} \;\dd z_2=\int_Y \,{\tau^0_{12}}_{|{z_1=-z_m}} \;\dd z_2$. Next, integrating over $Z$  the scalar   $\ba\cdot \div_\bz \btau^0=a_i \drs{\tau_{ij}^0}{z_j}=0$ (since $\div_\bz \btau^0=\bo$) with  $\ba=z_2\,\eu-z_1\ed$ and
integrating by parts, we get that
\beq
0=\int_{\partial Z}a_i \;\tau^0_{ij} n_{j} \;\dd s=-\int_Y \,z_2\,{\tau^0_{11}}_{|{z_1=-z_m}} \;\dd z_2-z_m\int_Y \,{\tau^0_{12}}_{|{z_1=-z_m}} \;\dd z_2=-\int_Y \,z_2\,{\tau^0_{11}}_{|{z_1=-z_m}} \;\dd z_2.,\label{een1}
\eeq
(the integral on $\partial Z$ reduces to that on  $Y$ at $z=-z_m$). In the limit $z_m\to \infty$, where $\int_Y z_2 \tau^0_{11}\to \mo{z_2\sa^0_{11}}(\haute,\bx')$, and accounting for $\sa^0_{11}$ in  \eqref{kim3}, we obtain the expected boundary condition
\beq
\drd{\Ua_2^0}{y_1}(\haute,\bx')=0.\label{ssx1}
\eeq
%
%
%

\subsection{Effective transmission conditions between the substrate and the region of the array}
\label{eff3}
To begin with, we shall need the solution in the substrate which is expanded  as
\beq\label{treza}
\bu=\sum_{n\geq0}\ep^n\bu^n(\bx), \quad \sig(\bx)=\sum_{n\geq0}\ep^{n}\sig^n(\bx), 
\eeq
with no dependence on the rapid coordinates, while in the array it is given by \eqref{exp}. 
As in the previous section,  a zoom is performed in the vicinity of the interface between  the substrate and the region of the array, owing to the substitution $y_1\to z_1$. 
In the intermediate region, the fields are expanded as in  \eqref{toto1} with the interface at $z_1=0$ and $z_1\in(-\infty,+\infty)$, see Figure \ref{Fig3}(b). It is worth noting that for $z_1\in(-\infty,0)$ the terms in the expansion \eqref{toto1} are assumed to be periodic with respect to $z_2\in (-\per/2,\per/2)$ while for $z_1\in(0,\infty)$ we have $z_2\in(-\cphi\per/2,\cphi\per/2)$.
Note that  we should use different notations for the expansions and for $z_1$ since their meaning is different;  for simplicity, we keep the same.  
The transmission conditions will be obtained by matching the solution in \eqref{toto1} for $z_1\to + \infty$ with that 
in \eqref{exp}  for  $x_1\to 0^+$, and  for $z_1\to - \infty$ with that 
in \eqref{treza} for  $x_1\to 0^-$. 
Matching the solutions hence means, with $\bz=(z_1,z_2)$, 
\beq\nonumber\toutin
\bv^0(\bz',\bx')+\ep \bv^1(\bz',\bx')+\cdots \underset{z_1 \to -\infty}{\sim}  \bu^0(\ep^2 z_1,\bx')+\ep \bu^1(\ep^2 z_1,\bx')+\cdots,
\\[10pt]
\bv^0(\bz',\bx')+\ep \bv^1(\bz',\bx')+\cdots \underset{z_1 \to +\infty}{\sim}  \bua^0(\ep z_1,z_2,\bx')+\ep \bua^1(\ep z_1,z_2,\bx')+\cdots,
\toutout
\eeq
where we have used that $x_1=\ep^2 z_1$ and $y_1=\ep z_1$. It results that 
\beq\label{mc1}\toutin
\dsp \lim_{z_1\to -\infty}\bv^0(\bz',\bx')= \bu^0(0^-,\bx'),\quad \lim_{z_1\to -\infty}\bv^1(\bz',x_2)= \bu^1(0^-,\bx'),\\
\dsp \lim_{z_1\to -\infty}\btau^0(\bz',\bx')= \sig^0(0^-,\bx'),\quad \lim_{z_1\to -\infty}\btau^1(\bz',\bx')= \sig^1(0^-,\bx'),
\toutout
\eeq
and that 
\beq\label{msc2}\toutind
\dsp \lim_{z_1\to +\infty}\bv^0(\bz',\bx')= \bua^0(0^+,\bx'),\quad &\dsp  \lim_{z_1\to +\infty}\left(\bv^1(\bz',\bx')-z_1\dr{\bua^0}{y_1}(0^+,z_2,\bx')\right)= \bua^1(0^+,z_2,\bx'),\\[10pt]
\dsp \lim_{z_1\to +\infty}\btau^0(\bz',\bx')= \bsa^0(0^+,z_2,\bx'), &\dsp \lim_{z_1\to +\infty}
\left(\btau^1(\bz',\bx')-z_1\dr{\bsa^0}{y_1}(0^+,z_2,\bx')\right)= \bsa^1(0^+,z_2,\bx').
\toutout
\eeq 
Eventually, with the differential operator in \eqref{gradz}, \eqref{eq11} applies; we shall also need  from \eqref{phys} that 
\beq\label{kim}
(\cc)\quad \ep^2 \sig=2\mu_\text{\tiny a} \,\ee^{\bz'}+\lambda_\text{\tiny a} \text{tr}(\ee^{\bz'}) \mI,\quad 
 \quad 
 (\cc')\quad \dsp \frac{1}{\ep}\ee^{y_1}+\ee^{\bx'}+\frac{1}{\ep^2} \ee^{\bz'}=\frac{(1+\nu_\text{\tiny a})}{E_\text{\tiny a}}\, \sig-\frac{\nu_\text{\tiny a}}{E_\text{\tiny a}}\,\text{tr}(\sig)\,\mI.
 \eeq
($\ee$ stands for $\ee(\bu)$) applying in the substrate, a=s, and in the plate, a=p, where we have defined
\beq
 \ee^{\bz'}(\bu)=\frac{1}{2}\begin{pmatrix} 
2\drs{u_1}{z_1}  & \left(\drs{u_1}{z_2}+\drs{u_2}{z_1}\right) & \drs{u_3}{z_1} \\[8pt]
\left(\drs{u_1}{z_2}+\drs{u_2}{z_1}\right) & 2\drs{u_2}{z_2} & \drs{u_3}{z_2} \\[8pt]
\drs{u_3}{z_1} & \drs{u_3}{z_2} & 0  
\end{pmatrix}.
 \eeq

The continuity of the displacement is easily deduced. 
From $(\cc')^{-2}$ in \eqref{kim},   $v_3^0$ does not depend on $\bz'$,  and $(v_1^0,v_2^0)$ correspond to a rigid body motion, {\em i.e.} $v^0_1=\Omega^0_\text{\tiny a} z_2+{V_1^0}_\text{\tiny a}$ and $v^0_2=\Omega^0_\text{\tiny a} z_1+{V_2^0}_\text{\tiny a}$, with $(\Omega^0_\text{\tiny a},{\bf V}^0_\text{\tiny a})$ independent of $\bz'$. The periodic boundary conditions in the substrate impose $\Omega^0_\text{\tiny s}=0$; next, the continuity of the displacement at $z_1=0$ imposes $\Omega^0_\text{\tiny p}=0$ in the plates and $\bv^0={\bf V}^0_\text{\tiny p}={\bf V}^0_\text{\tiny s}$ is independent of $\bz'$. From \eqref{mc1}-\eqref{msc2},  $\bu^0(0^-,\bx')=\bv^0=\bua^0(0^+,z_2,\bx')$, and making use of \eqref{kim2}
\beq\label{trtr}
u^0_1(0^-,\bx')=\Ua_1^0(\bx'),\quad u^0_2(0^-,\bx')=\Ua_2^0(0^+,\bx').
\eeq
For the same reasons, $\bv^1$ is  a constant displacement, hence 
$\bu^1(0^-,\bx')=\bua^1(0^+,z_2,\bx')$, but this has now a consequence. Indeed, 
from  \eqref{msc2}  for the displacement at order 1, we have necessarily  $\partial_{y_1}{\bua^0}(0^+,z_2,\bx')=\bo$ to ensure that $\bua^1(0^+,z_2,\bx')$ is finite;  from \eqref{kim2}, we already know that  $\drs{\ua^0_1}{y_1}=\drs{\ua^0_3}{y_1}=0$ but the condition remains for $\ua^0_2=W^0_2(y_1,\bx')$, hence 
\beq
 \dr{\Ua^0_2}{y_1}(0^+,\bx')=0.\label{ssj1}
\eeq

We now move on the effective conditions on the force. From $(\cc)^{-2}$  in \eqref{eq11},   
$\div_{\bz'} \btau^0=\bo$ that we integrate   over $Z=\{z_1\in(0,z_m), z_2\in Y\}\cup\{z_1\in(-z_m,0), z_2\in(-\per/2,\per/2)\}$. Accounting for i)  $\btau^0\cdot {\bf n}$  continuous at $z_1=0$, ii)  $\btau^0\cdot {\bf n}=0$ between the plates and the air, iii)  $\btau^0$  periodic at $z_2=\pm\per/2$ in the substrate, 
 we get that  $ \int_{-\per/2}^{\per/2}\tau_{1i}^0(-z_m,z_2,\bx')\,\dz_2= \int_Y \tau_{1i}^0(z_m,z_2,\bx')\,\dz_2$. In the limit $z_m\to \infty$ in  \eqref{mc1} - \eqref{msc2} along with  $\mo{\sa^0_{1i}}=0$ from \eqref{kim1}, we get 
 \beq
 \dsp    \sigma^0_{1i}(0^-,\bx')=0, \quad i=1,2,3, 
 \label{yuyu0}
 \eeq
which tells us that 
 the plates do not couple to the substrate at the dominant order.  The coupling appears at the next order, starting with  $\div_{\bz'} \btau^1=\bo$ from $(\cc)^{-1}$. As for $\btau^0$ and using again that $\mo{\sa^0_{1i}}=0$, we get that      $\sigma^1_{1i}(0^-,\bx')=\mo{\sa^1_{1i}}(0^+,\bx')$; eventually, using $\mo{\sa^1_{1i}}$ in \eqref{kim4}, we get 
\beq\label{yuyu}
\toutin
\dsp \sigma_{11}^1(0^-,\bx')= \rt\omega^2\cphi\haute\,\Ua_1^0(\bx'),\quad \sigma_{12}^1(0^-,\bx')=-\;\frac{\Et}{(1-\nut^2)}\,\frac{\cphi^3\per^2}{12}\, \drt{\Ua^0_2}{y_1}(0^+,\bx'),\\
\dsp  \sigma_{13}^1(0^-,\bx')=\cphi\haute \left(\Et\drd{W^0_3}{x_3}(\bx')+\rt\omega^2 W^0_3(\bx')\right)
\toutout
\eeq
   
\subsection{The final problem}
\label{effin}
The effective problem \eqref{CL} is obtained for $(\bu=\bu^0,\sig=\sig^0+\ep\sig^1)$ in the substrate for $x_1<0$, $(\bu={\bf W}^0,\sig=\bsa^0+\ep\bsa^1)$ in the region of the array for $x_1>0$. 
Remembering that   $y_1=x_1/\ep$ and $\haute=\haut/\ep$, $\per=\Per/\ep^2$, it is easy to see that (i) the Euler-Bernoulli equation in \eqref{CL} is obtained from \eqref{kim5}, (ii) the effective boundary conditions announced in \eqref{CT1} from \eqref{jwx3}, \eqref{ssx1}, \eqref{trtr}, \eqref{ssj1} and   \eqref{yuyu0}-\eqref{yuyu}.


\section{Numerical validation of the effective problem for a two-dimensional problem} 
\label{secnum}

In this section, we inspect the validity  of the effective problem in a two-dimensional setting for in-plane waves ($u_3=0$, hence $\partial_{x_3}=0$). We solve numerically the actual problem of an incident plane wave coming from $x_1\to -\infty$ at oblique incidence on the free surface supporting the array of plates, and Lamb waves are excited in the plates. This is done using a multimodal method with pseudo-periodic solutions in the soil and Lamb modes in the plates; the method is detailed in \cite{petitmonstre}. 
In the effective problem, the solution is explicit, from \eqref{CL} - \eqref{CT1} or equivalently  \eqref{CT2}-\eqref{deff} when the solution in the plates is disregarded.

We set the  material properties for the elastic substrate: $\nug=0.2$, $\Eg=2$ GPa, $\rg=1000$ Kg.m$^{-3}$,
and for the plates : $\nut=0.3$, $\Et=2$ GPa, $\rt=500$ Kg.m$^{-3}$, and $\cphi=0.5$.
We choose $\Per=1$ m  and we set $\ep=\sqrt{\kkt \Per}=0.37$ ($\omega=124$ rad.s$^{-1}$), hence $\kk=0.64$ m$^{-1}$. 
We shall consider $\haut\in(0,30)$ m resulting in $\kk \haut\in(0,20)$ which includes the first 6 bending modes for $h=h_n$, $n=1,\cdots,6$, and $h_1\simeq 3$ m, $h_2\simeq 7.3$ m, $h_3\simeq 12.3$ m, $h_4=17.2$ m,  $h_5=22.1$ m, $h_6=27.0$ m. The first resonance of the quasi-longitudinal wave along $x_1$ appears for $\haut=\pi/(2\omega)\; \sqrt{\Et/\rt}\simeq 25.3$ m, hence it will be visible in our results.

\subsection{Reflection of elastic waves - the solution of the effective problem}
 
 \begin{figure}[b!]
\centering
\includegraphics[width=.65\columnwidth]{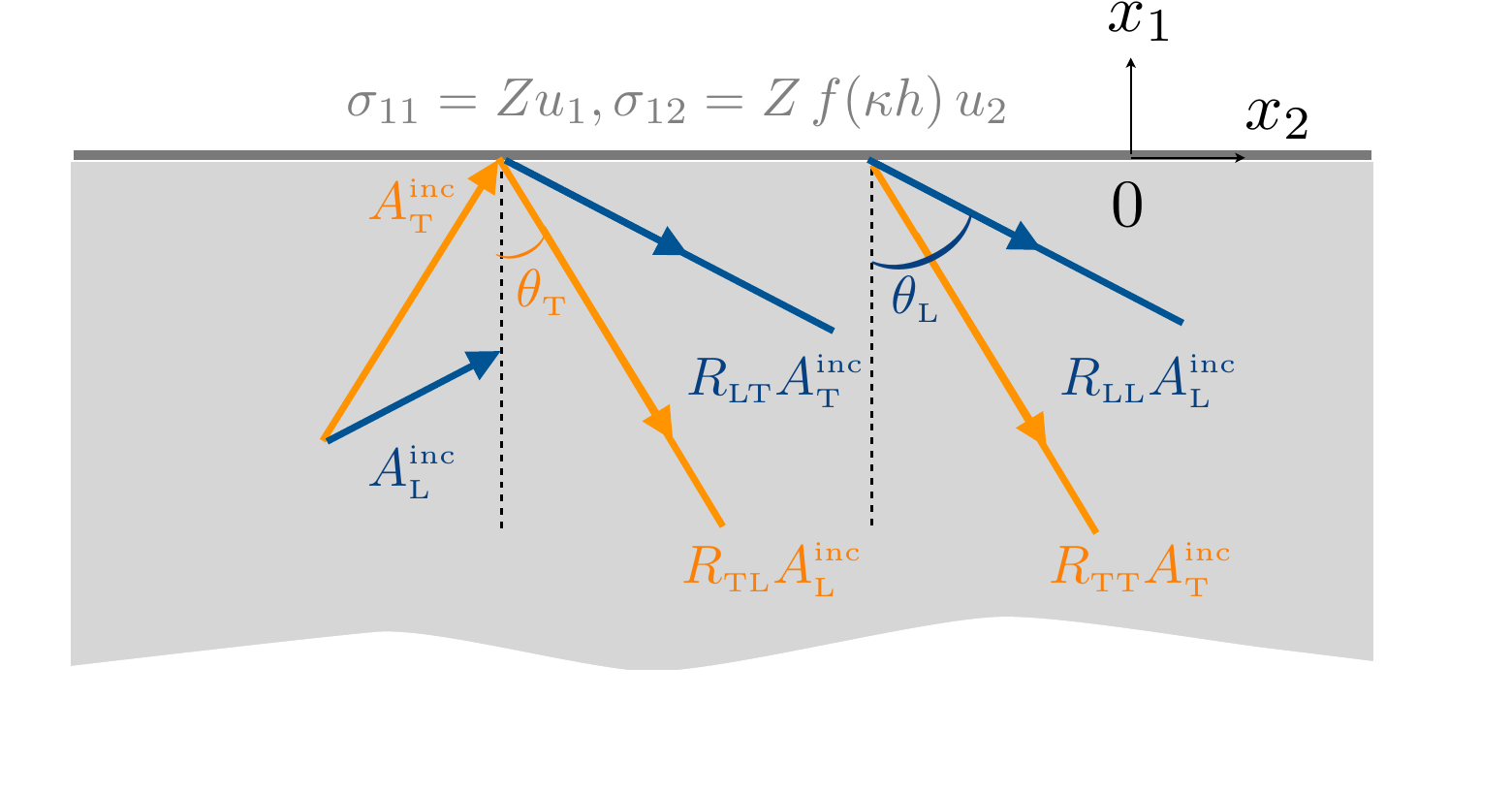}
	\caption{The effective (partial) problem -- Reflection of an elastic wave on the surface $x_1=0$ where effective boundary conditions \eqref{CT2} apply. The incident wave is defined by   \eqref{incident} and the solution by \eqref{phipsi}. The arrows show the wavevectors.}
	\label{fig9}
\end{figure}
We define the potentials  $(\phi,\psi)$ using the Helmholtz decomposition, with  $\bu=\grad \phi+\grad \times (\psi\; {\bf e}_3)$.
The incident wave in the substrate is defined in terms of the incident potentials 
\beq\label{incident}
\toutin
\dsp \phii(x_1,x_2)=\aph e^{i\kl x_1} e^{i\be x_2}, \quad \psii(x_1,x_2)= \aps e^{i\kt x_1}e^{i\be x_2},\\[10pt]
\dsp \text{with }
(\kl,\be)=\kkl\;(\cos\tl,\sin\tl), \quad (\kt,\be)=\kkt\;(\cos\tt,\sin\tt),
\toutout\eeq
with $\kkl=\sqrt{\frac{\rg}{\lamg+2\mug}}\;\omega$ and $\kkt=\sqrt{\frac{\rg}{\mug}}\;\omega$.
The solution in the substrate reads   
\beq\label{phipsi}
\toutin
\dsp \phi(x_1,x_2)=\phii(x_1,x_2)+\left(\rll \,\aph+\rlt \,\aps\right) e^{-i\kl x_1} e^{i\be x_2}, 
\\[10pt]
\psi(x_1,x_2)=\psii(x_1,x_2)+\left(\rtl \,\aph+\rtt \,\aps\right) e^{-i\kt x_1} e^{i\be x_2}.
\toutout\eeq
The effective problem can be solved explicitly. Accounting for the boundary conditions \eqref{CT2},  it is easy to show that 
\beq\label{lesR}
\toutin
\dsp \rll=\frac{1}{D}\;\left[\sin2\tt \sin2\tl-\kn^2\cos^22\tt-i\kn\au \left( \cos\tl-\kn \ff\cos\tt \right)-\xi\au^2\ff \cos(\tl+\tt)\right], \\[12pt]

 \dsp \rtt=\frac{1}{D}\;\left[\sin2\tt \sin2\tl-\kn^2\cos^22\tt+i\kn\au\left(\cos\tl-\kn\ff \cos\tt\right)-\xi\au^2\ff\cos(\tl+\tt)\right], \\[12pt]

\dsp \rlt=-\frac{\kn^{2}\sin2\tt}{D}{\,\left(2\cos2\tt+\au^2\ff\right)}, \quad

\dsp \rtl=\frac{\sin2\tl}{D}\;{\left(2\cos2\tt+\au^2\ff\right)}, \\[10pt]
\begin{array}{ll}
\text{where } 
& \dsp D=\sin2\tt \sin2\tl+\kn^2\cos^22\tt-i\kn\au\left( \cos\tl+\kn\ff \cos\tt\right)-\xi\au^2\ff\cos(\tl-\tt),\\[8pt]
\dsp
& \dsp \text{$\ff$ stands for $\ff(\kk\haut)$  in \eqref{deff} and  } \kn=\frac{\sin\tl}{\sin \tt}=\sqrt{\frac{2(1-\nug)}{1-2\nug}},\quad \au=\cphi\,\frac{\rt}{\rg}\;\kkt \haut, 
\end{array}
\toutout
\eeq
(note that $\au=\epun/(\kkt \rg)$ in \eqref{deff}).  

 \begin{figure}[h!]
\centering
\includegraphics[width=.65\columnwidth]{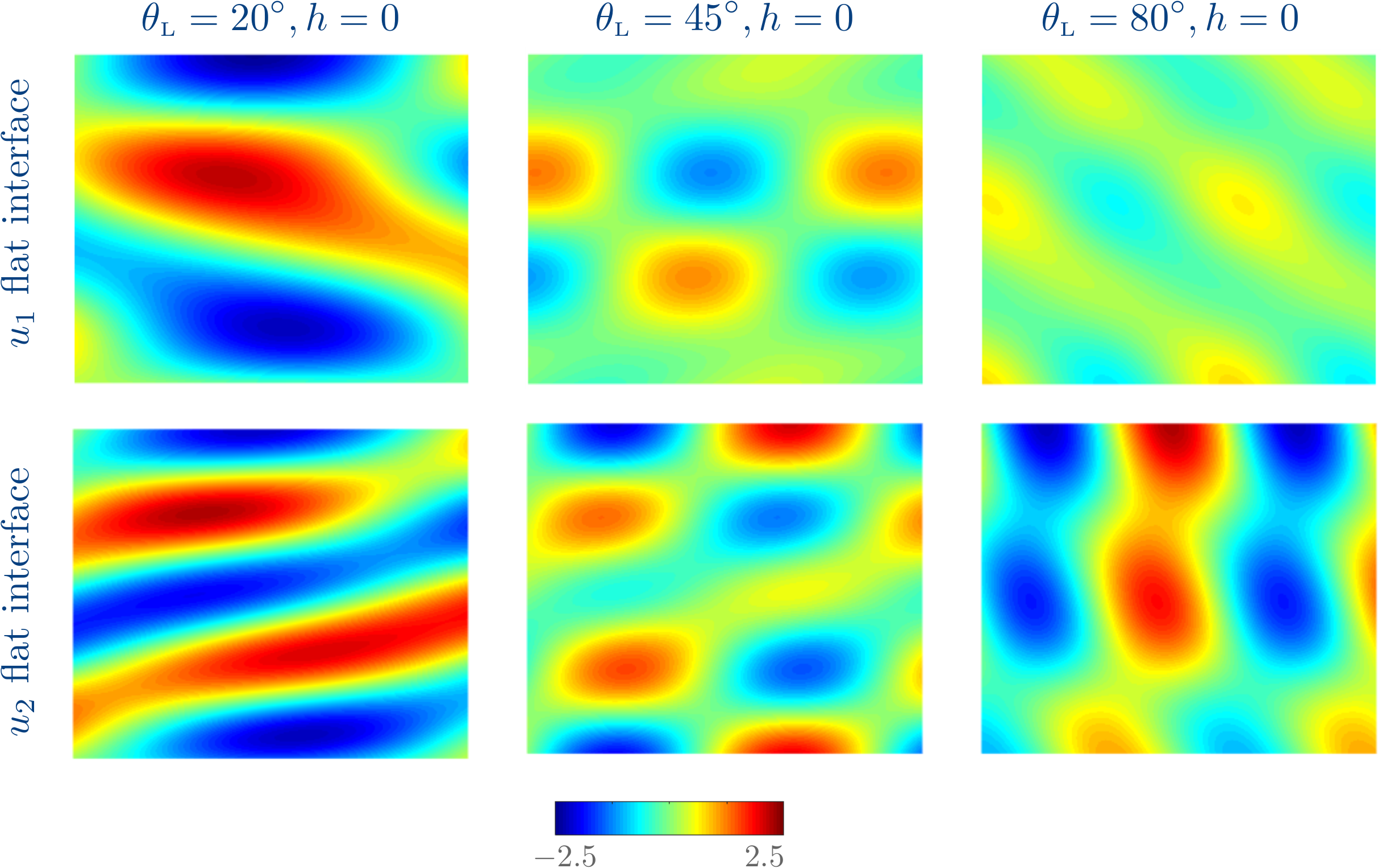}
	\caption{Reference case -- Vertical 
	$u_1$ and horizontal $u_2$ displacement fields in the absence of array. The fields are shown for $x_1\in (-100,0)$ m and $x_2\in (0,120)$ m.}
	\label{fig4}
\end{figure}

Obviously, the same reflection coefficients are obtained by solving the complete problem \eqref{CL}-\eqref{CT1}; we get the displacement fields in the region of the plates $x_1\in(0,\haut)$, with 
\beq\label{u12}
\begin{array}{c}
\dsp u_1(x_1,x_2)=u_1(0,x_2),\quad {u_2}(x_1,x_2)=u_2(0,x_2)V(x_1),
\\[8pt]
\toutin
\dsp u_1(0,x_2)=\frac{2i\kn^2 \kkl}{D} \left[
\cos\tl\left(\cos 2\tt-i\cos\tt \;\au \ff\right)\aph +\left(\sin 2\tt\cos\tl-i\sin\tl\cos\tt \;\au \ff\right)\aps\right]
\;e^{i\beta x_2},\\[14pt]
 \dsp u_{2}(0,x_{2})=\dsp 
 \frac{i\kkt}{D}\; \left[\sin2\tl (2\cos\tt - i\au)\aph-2\kn \cos\tt(\kn \cos 2\tt-i\au\cos\tl )\aps\right]\;e^{i\beta x_2}, \\[14pt]
\dsp V(x_1)=\frac{v_1(\kk\haut)\left[\ch \kk(x_1-\haut)+\cos \kk(x_1-\haut)\right]+
v_2(\kk\haut)\left[\sh \kk(x_1-\haut)+\sin \kk(x_1-\haut)\right]}{2\left(1+\ch\kk\haut\cos\kk\haut\right)},\\[10pt]
\text{with  }\quad  v_1(\kk\haut)=(\ch \kk\haut+\cos \kk\haut), \quad v_2(\kk\haut)=(\sh \kk\haut-\sin \kk\haut).
\toutout 
\end{array}\eeq

As a reference case, typical  displacement fields $(u_1,u_2)$ for a surface on its own ($\haut=0$) are reported in figure \ref{fig4}. The  incident wave is of the form  \eqref{incident}  with $\aph=1/(2\beta), \aps=-1/(2\kt)$ producing an incident horizontal displacement equal to unity at $x_1=0$; three incident angles  $\tl$ are reported. It is worth noting that with  $\au=0$  in \eqref{lesR}, we recover the reflection coefficients for  a flat interface, see {\em e.g.} \cite{achenbach}.

\subsection{Weak and strong interactions between plates and substrate}

\begin{figure}[h!]
\centering
\includegraphics[width=.6\columnwidth]{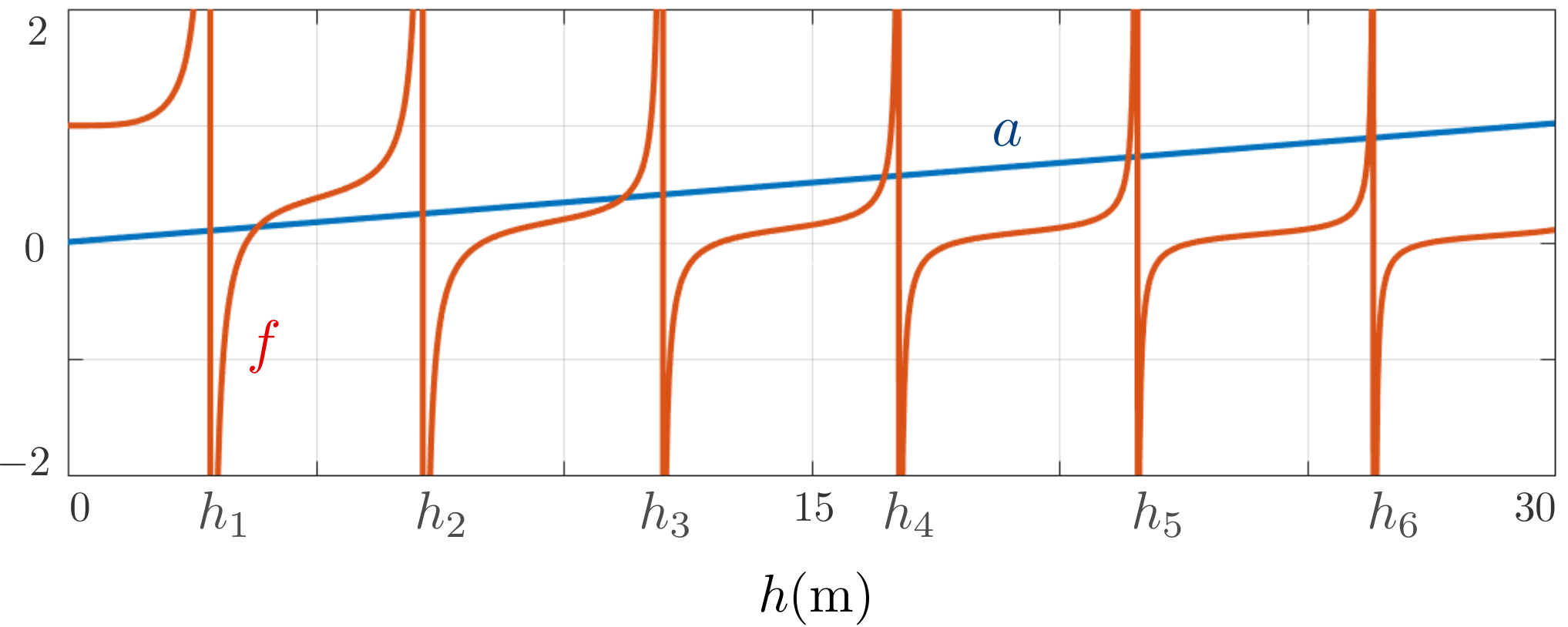}
	\caption{Variations of the impedance parameters $\au=Z/(\kkt \rg)$ in \eqref{lesR}, and $(Z,\ff)$ in \eqref{deff}, versus $\haut$. The parameter $\ff$ diverges at the resonances of the clamped - free plates for $\ch\kk \haut \cos \kk \haut=-1$.
	}
	\label{fignn01}
\end{figure}

The effect  of the array is encapsulated in the impedance parameters  $(\au,\ff)$, or equivalently  $(\epun=\kkt \rg\au,\ff)$, whose  variations  versus $\kkt \haut$ are reported in figure \ref{fignn01}. The parameter $\epun=\rt\omega^2\cphi\haut$ tells us that  heavier  plates   and higher frequency produce  more pronounced coupling with the substrate, which is not surprising. The pa\-ra\-me\-ter  $\ff$ encapsulates the effects of the bending resonances and it diverges when approaching them.  This occurs at the frequencies corresponding to a clamped- free  single plate, in other words $\ch\kk \haut\cos \kk \haut=-1$.  

It is worth noting that the impedance condition implying the resonant contribution is used in \cite{boutin2013}
in the same configuration considering  a spring-mass model. Neglecting the damping used in this reference and adapting the notation, the impedance parameter $f$ is reduced to $f=1/(1-\omega^2/\omega_n^2)$ with $\omega_n$ a resonance frequency. This relation has to be understood locally in the vicinity of a single resonance, and  from $f$ in figure \ref{fignn01}, it is visible that (i) it captures the  physics of a single resonance locally (ii) it should be corrected by a  constant $C_n$ depending on the considered resonance  $f=C_n/(1-\omega^2/\omega_n^2)$ to account for the increase in sharpness of the bending resonances when $h$ or $\omega$ increases.

\subsubsection{Weak interaction}

\begin{figure}[h!]
\centering
\includegraphics[width=.9\columnwidth]{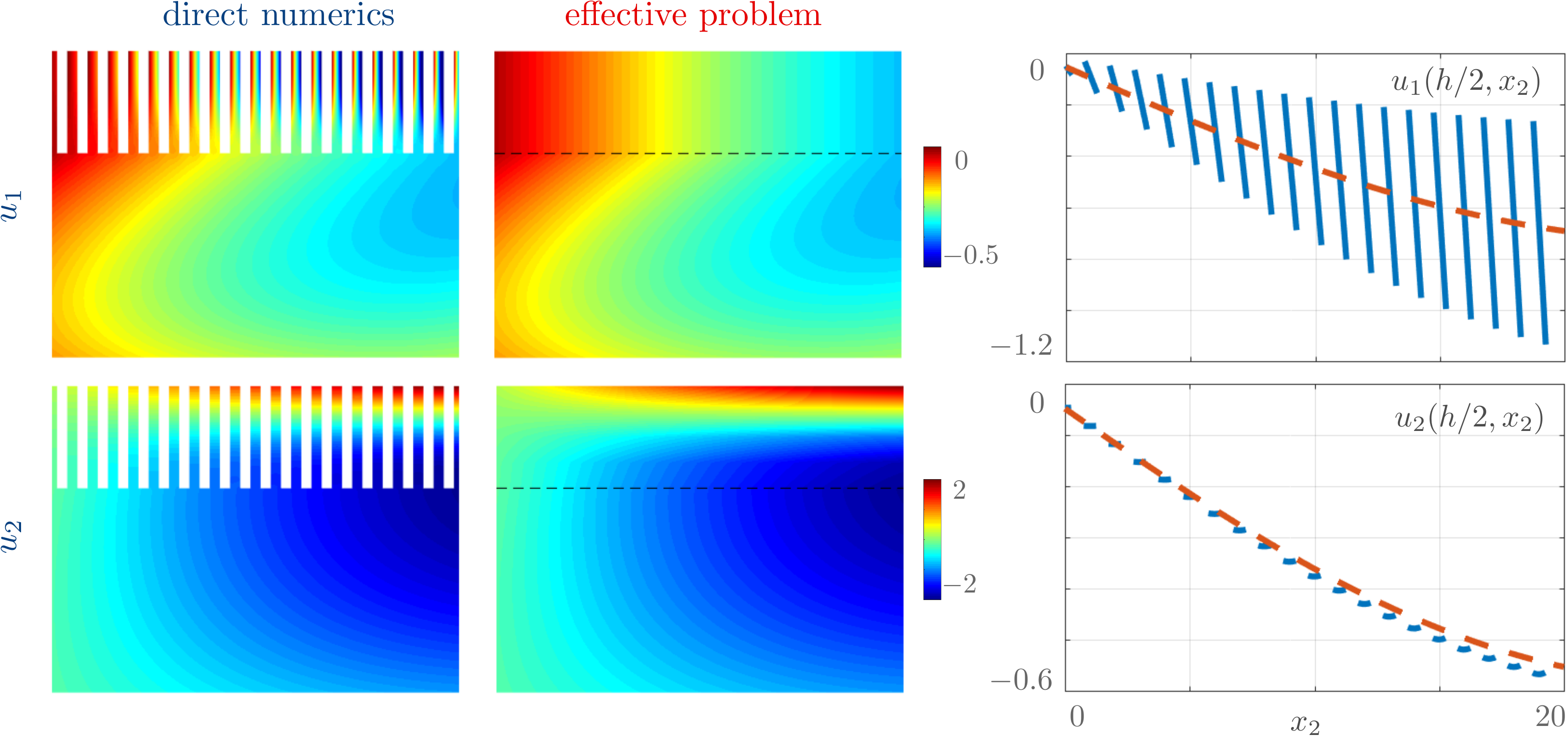}
	\caption{Meaning of the homogenization process -- 
	Displacement fields (actual and homogenized)  for $x_1\in (-10,\;5)$ m and $x_2\in (0, \;20)$ m ($\haut=5$ m, $\tl=45^\circ$).}
	\label{FigZoom}
\end{figure}

Far from the resonances,  the interaction is weak. Indeed, from \eqref{CT2}, with  $\epun$ being small and $f(\kk\haut)$ of the order of unity, the wave impinging the surface sees essentially a flat surface, with $\sigma_{11}\simeq \sigma_{12}\simeq 0$ at $x_1=0$. The  resulting patterns, not reported,
 are indeed similar to those obtained for a flat interface in figure \ref{fig4}. Since there is not  much to be said on the field in the substrate,  we focus on the capability of the complete effective solution to reproduce the actual displacement in the plates.   
Figure \ref{FigZoom} show a small region of the displacement fields near the interface ($\haut=5$ m resulting in $\kkt \haut=0.7$ and $\tl=45^\circ$). From what we have said (the interaction is weak), 
the displacements in the substrate are neatly reproduced. More interestingly, the displacements in the plates are also accurately reproduced in an "averaged" sense which clearly appears for the displacement $u_1$ : in the actual problem, $u_1$ varies linearly with $x_2$ within a single plate, 
in agreement with  \eqref{inter1}; this variation at the small scale is superimposed to a variation
at large scale, from one plate to the other.  
The small scale variations do not appear in the homogenized solution since they vanish on average while
 the large scale variation is  captured. 
 The same occurs for $u_2$ but in this case, 
 the small scale   variations  are less visible 
  because they appears at the order 2 (see \eqref{kim2} and \eqref{w22}).  

\subsubsection{Strong coupling near the resonances}
\begin{figure}[h!]
\centering
\includegraphics[width=.85\columnwidth]{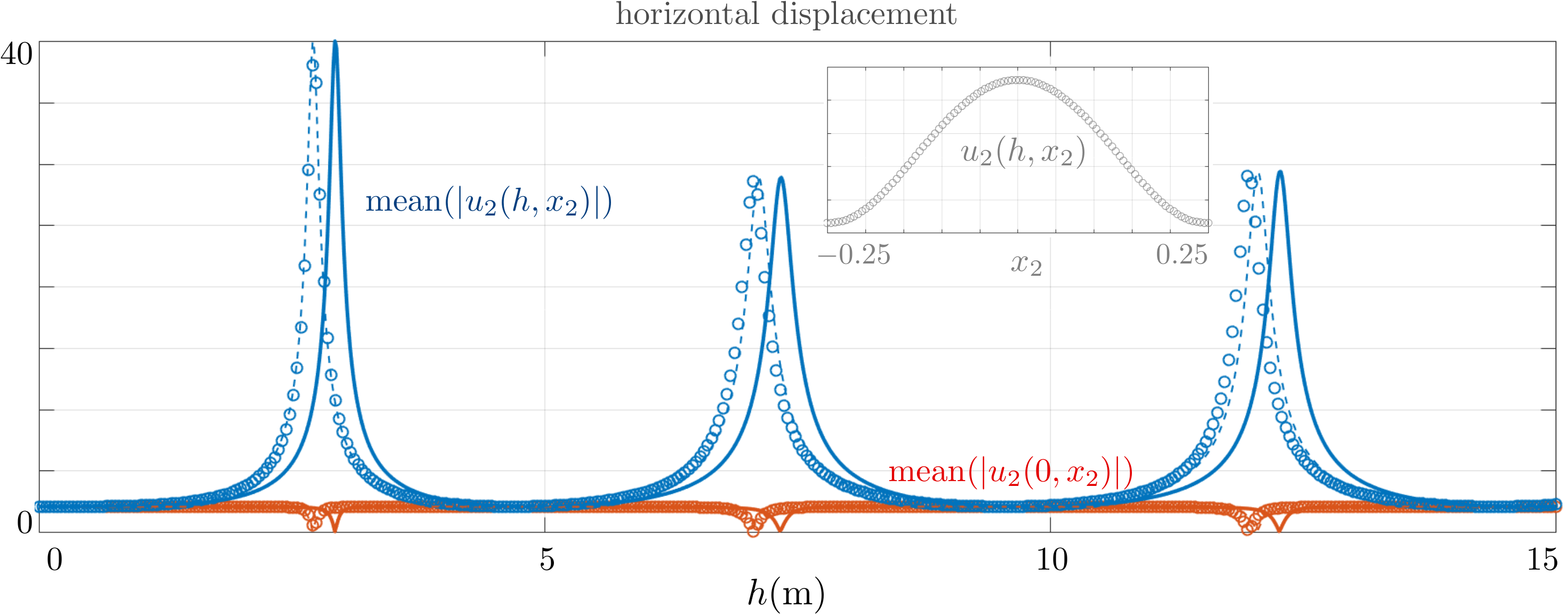}
	\caption{Amplitude of  $u_2$ at the bottom $x_1=0$ and at the top $x_1=\haut$ of a  plate against $\haut$, from direct numerics (symbols) and from the effective solution \eqref{u12} (plain line); $\tl=45^\circ$. Shifts in the resonances $\haut\to \haut-h_0$, with $h_0=0.22$ m are compensated (dashed line). The inset shows the actual variation of $u_2(\haut,x_2)$ within a single plate with variations as small as about $10^{-4}$ with respect to the mean value. }
	\label{figres}
\end{figure}

Strong coupling in the vicinity of the bending resonances can be measured by the  amplitudes 
of the displacements in the plates. 
We report in figure  \ref{figres} the amplitudes of the horizontal
 displacements against $\haut$, at the bottom and at the top of a single plate. 
 In the actual problem these amplitudes  are calculated by averaging 
 over $x_2\in(-\htt/2,\htt/2)$ the profiles $|u_2(0,x_2)|$ and $|u_2(\haut,x_2)|$  obtained numerically.
  In the homogenized problem   
  $|u_2(0,x_2)|$ and $|u_2(\haut,x_2)|=|V(\haut)u_2(0,x_2)|$ are given in 
  closed-forms from  \eqref{u12}. 
  
For $\haut\in (0, 15)$ m, the first three bending resonances are visible by means of high displacements at the top of the plates (up to 40 times the amplitude of   the  incident wave in the reported case). It is also visible by means of vanishing amplitude at the bottom of the plate, in agreement with  \eqref{u12} for $\ff\to\infty$. Hence, near the bending resonances, the plates are clamped and they impose a vanishing horizontal displacement at the interface with the substrate, a fact already mentioned in \cite{boutin2013}.
 \begin{figure}[b!]
\centering
\includegraphics[width=.7\columnwidth]{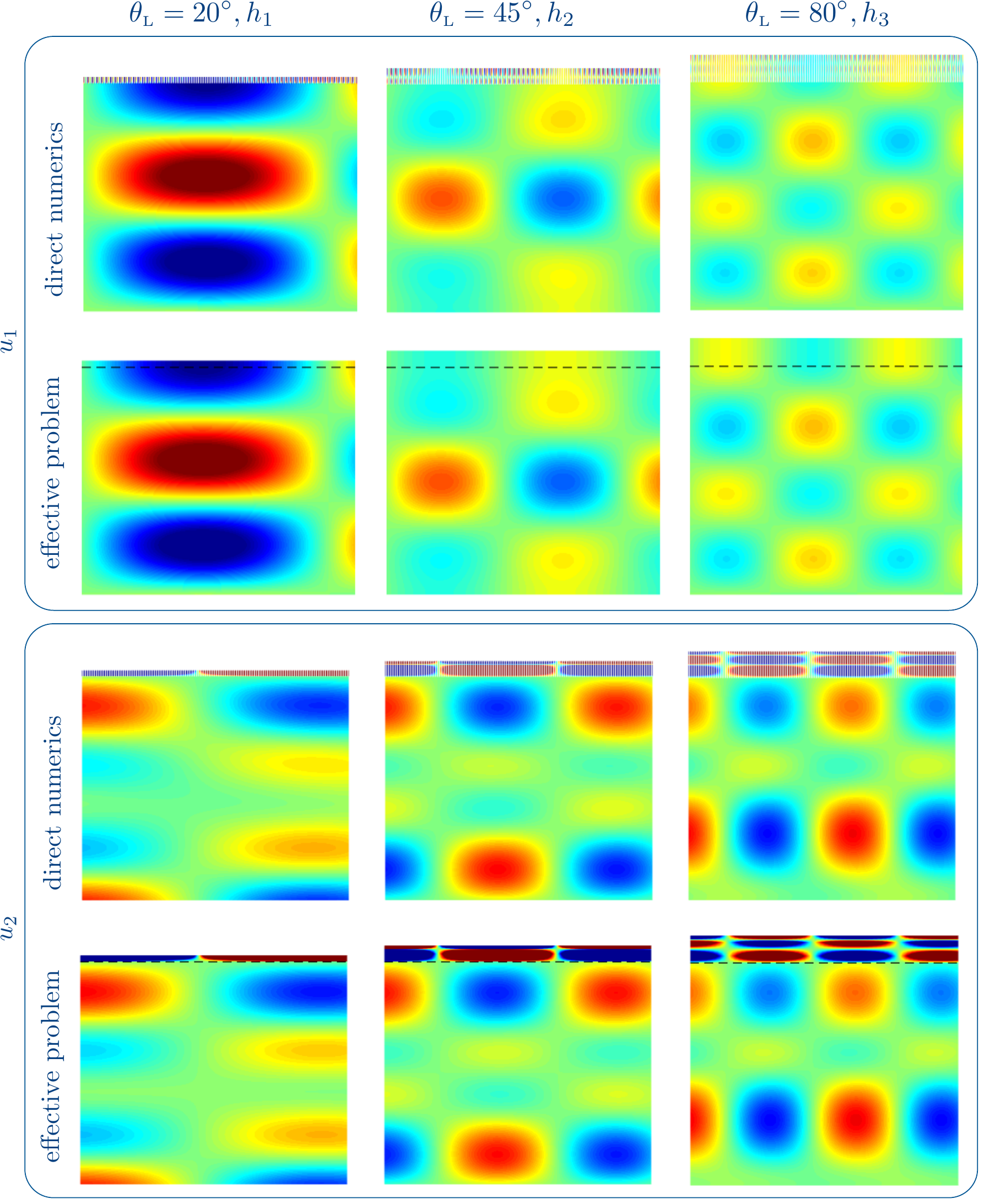}
	\caption{Strong coupling of the array near the three first bending resonances -- The values of $h_n$, $n=1,2,3$ have been taken from figure \ref{figres} at the maximum displacements in the direct numerics and in the homogenized solution. 	Same representation as in figure \ref{fig4}.   }
	\label{fig8}
\end{figure}
In the substrate, the resulting displacements are significantly impacted. Large values of $\ff(\kk \haut)$ 
produce $\rll\simeq -1$, $\rtt\simeq 1$ and $\rtl\simeq \rlt\simeq 0$ in \eqref{lesR}, hence
\beq
\toutin
\dsp u_1(x_1,x_2)\simeq 2i \kkl \cos\tl \left( \aph\cos \kl x_1 +\aps \tan\tl \cos\kt x_1\right)e^{i\be x_2},\\[12pt]
\dsp u_2(x_1,x_2)\simeq 2 \kkt \cos\tt \left( -\aph\tan\tl \sin \kl x_1+ \aps\sin\kt x_1 \right)e^{i\be x_2},
\toutout
\eeq 
corresponding to a superposition of standing waves. 
 Examples of resulting patterns are shown  in figure \ref{fig8} for the  first three bending resonances to be compared with those obtained for a flat interface in figure \ref{fig4}. 
 It is worth noting that in figure \ref{fig8} we have accounted for the shift in $h_n$, $n=1,2,3$ in the homogenized solution from figure \ref{figres}, where  a systematic shift of the effective solution $\haut\to \haut-h_0$, with $h_0=0.22$ m in the present case.

\subsection{Occurence of the first longitudinal resonance}

\begin{figure}[h]
\centering
\includegraphics[width=1\columnwidth]{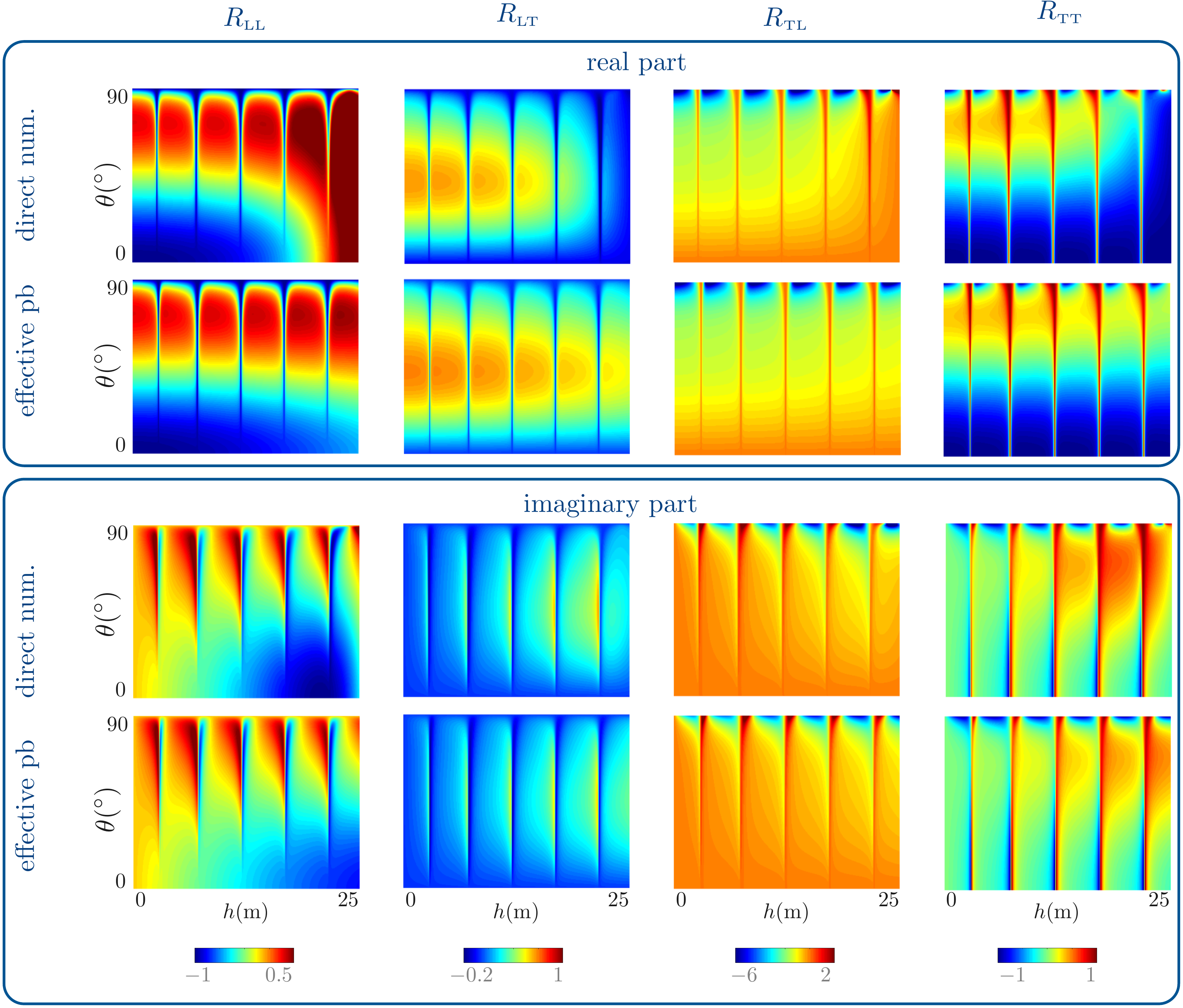}
	\caption{Accuracy of the effective model on the reflection coefficients -- Real and imaginary parts of the four reflection coefficients versus $h$ and $\theta$. The solution of the effective problem, \eqref{lesR} (second row) reproduces accurately the direct numerical solution of the actual problem (first row) up to the occurence of the first longitudinal resonance  for $h\simeq 25.3$ m.}
	\label{Fig10}
\end{figure}

To go further in the analysis, we report in figure \ref{Fig10} the reflection coefficients against $h\in(0,25)$ m and $\theta\in(0,90^\circ)$. We represent  the real and imaginary parts of the 4 reflection coe\-ffi\-cients. 
As previously said, our analysis does not hold at and  above the first longitudinal re\-so\-nance, 
which appears  for $\haut\simeq 25.3$ m (hence $\kk\haut=3.4$); expectedly, the effective model indeed breaks down at this high value but it remains surprisingly  accurate up to $h\sim 15$ m (hence $\kkt \haut\sim 2$).

\begin{figure}[h]
\centering
\includegraphics[width=.8\columnwidth]{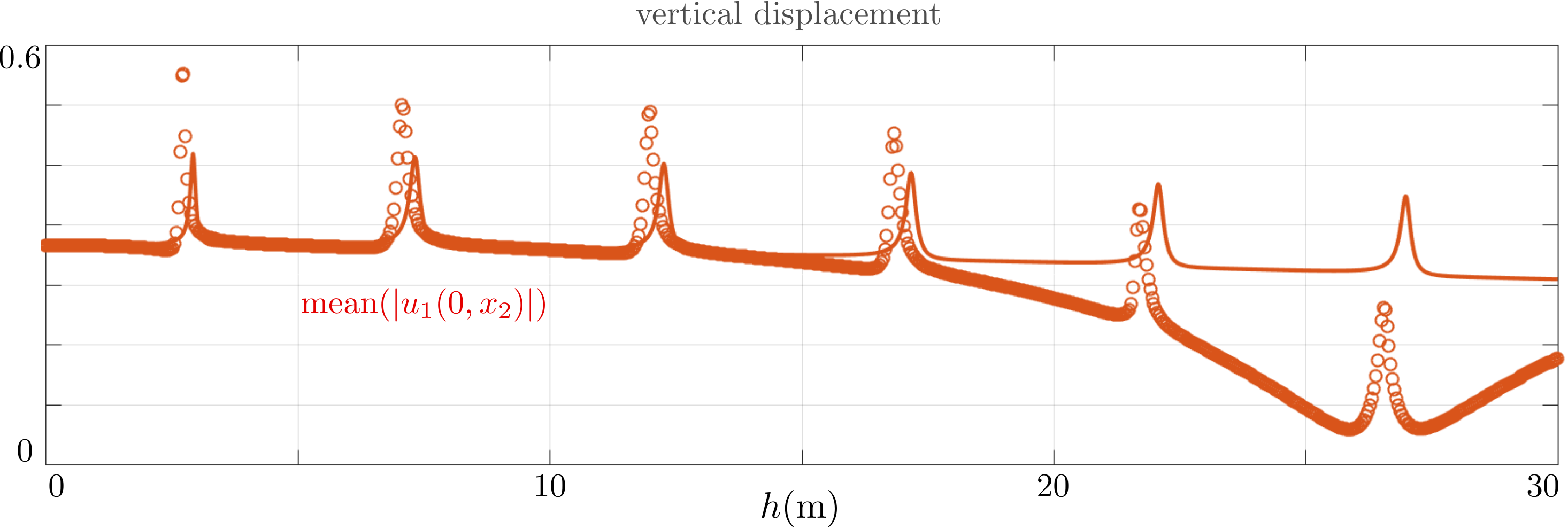}
	\caption{Occurence of the first resonance in reflection -- Variation of the amplitude of the vertical displacement $u_1(0,x_2)$ at the bottom of the plates. The bending resonances are superimposed to this resonance which produces a almost clamped condition $u_1(0,x_2)=0$ for $h\simeq 25.3$ m. }
	\label{Figlong}
\end{figure}

The occurence of this resonance is visible by means of the amplitude of the vertical displacement $u_1(0,x_2)$, which is reported in figure \ref{Figlong} against $\haut$. We observe the same trends  as for the bending modes. Far from the resonance, the displacement is essentially the same as for a  surface on its own; at the longitudinal resonance, it tends to zero resulting in  clamped- free conditions for the plates.  However, it is also visible that rapid variations of the displacements due to multiple bending resonances superimpose to the smooth variations of the displacement due to the longitudinal resonance. 

\section{Conclusion}
\label{conclude}

We have studied the interaction of an array of plates or beams with an elastic half-space u\-sing asymptotic analysis and homogenization techniques. The resulting models \eqref{CL}-\eqref{CT1} for plates and \eqref{complet3}-\eqref{CT3d} for beams provide one-dimensional propagation problems which in their simpler form consist in effective boundary conditions at the surface of the ground,  \eqref{CT2} for plates and \eqref{CT3p} for beams. 
The exception for plates in the boundary condition $\sigma_{13}$ in \eqref{CT1} is  incidental for in-plane incidence 
but it is interesting since it provides non trivial coupling for arbitrary incidence. 
For in-plane incidence, the model  has been validated by comparison with direct numerical simulations which show an  overall good agreement. In particular, the   displacement fields obtained in a closed-form accurately reproduce the actual ones; this is of practical importance for applications to site-city interaction where the  displacements at the bottom and at the top of buildings are relevant quantities to measure the risk of building damage.  

\vspace{.3cm}
Our models have been obtained owing to a deductive approach which applies to a wide va\-rie\-ty of problems. An important point is that the analysis does not assume a preliminary model reduction for the resonator on its own and as such, it can be conducted at any order. Higher order models   would involve  enriched transmission and boundary conditions able to capture   more subtle effects as the shift in the resonance frequencies visible in the figure  \ref{figres} or the presence of heterogeneity at the roots and at the top of the bodies as it has been done in \cite{nousLove}.
Next, we have considered bodies with sufficient symmetry resulting in a diagonal rigidity matrices and which allow for easier calculations. When symmetries are lost, and the simplest case is that of beams with rectangular cross-sections, the calculations are similar; they will produce couplings for incidences as soon as the horizontal component does not coincide with one of the two principal directions.
Additional complexities can be  accounted for straightforwardly, as orthotropic anisotropy  along $x_1$ or slow variations in the cross-section.      
Eventually, the models are restricted to the  low frequency regime where only the flexural resonances take place. At the threshold of the first longitudinal resonance, they fail as illustrated in figure \ref{Figlong}. Extension of the present work consists in adapting the homogenization procedure in order to capture both flexural and longitudinal resonances at higher frequencies.

\vspace{1cm}

\noindent {\bf Acknowledgements}

A. M. and S.G. acknowledge insightful discussions with Philippe Roux at the Institute ISTerre of the University of Grenoble-Alpes. S.G. is also thankful for a visiting position in the group of Richard Craster within the Department of Mathematics at imperial College London in 2018-2019.

\appendix

\section{Remark on the solution in the region of the plates }
\label{aparbre}

From the boundary conditions  \eqref{CT1},  $\drd{u_2}{x_1}(\haut,x_2)=\drt{u_2}{x_1}(\haut,x_2)=0$ and  ${u_2}(0^+,x_2)={u_2}(0^-,x_2)$, $\dr{u_2}{x_1}(0^+,x_2)=0$, the general solution for $x_1\in(0,\haut)$ reads as follows
\beq\toutin
\dsp u_{2}(x_{1},x_2)=A(x_{2})\left\{
a(\kk\haut)\left[\ch \kk(x_1-\haut)+\cos \kk(x_1-\haut)\right]+
b(\kk\haut)\left[\sh \kk(x_1-\haut)+\sin \kk(x_1-\haut)\right]
\right\},\\[10pt]
\text{with  }\quad  a(\kk\haut)=(\ch \kk\haut+\cos \kk\haut), \quad b(\kk\haut)=(\sh \kk\haut-\sin \kk\haut).
\toutout\eeq
 The displacement $u_{2}$ is continuous at $x_1=0$ and we have  
$2A(x_{2})(1+\ch\kk\haut\,\cos\kk\haut)=u_2(0,x_{2})$,
hence
 \beq \toutin
 {u_2}(x_1,x_2)=u_2(0,x_2)V(x_1),\\[10pt]
\dsp V(x_1)=\frac{a(\kk\haut)\left[\ch \kk(x_1-\haut)+\cos \kk(x_1-\haut)\right]+
b(\kk\haut)\left[\sh \kk(x_1-\haut)+\sin \kk(x_1-\haut)\right]}{2\left(1+\ch\kk\haut\cos\kk\haut\right)}.
\toutout
\eeq

\vspace{.2cm}
Obviously, this holds except at the resonance frequencies of the plates for  $\ch\kk_{r}\haut\,\cos\kk_{r}\haut=-1$
which imposes $u_{2}(0,x_2)=0$ (and eigenmodes in the region of the plates). It follows that the relation on $ \sigma_{12}(0^-,\bx')$ in \eqref{CT1} becomes $ \sigma_{12}(0^-,\bx')=-(\Ben/\Per)V'''(0) \;u_2(0,x_2)$, with $V'''(0)=-\kk^4 \haut f(\kk\haut)$,  with  $f(\kk\haut)$ in \eqref{CT2}.
With  $\sigma_{12}(0^-,\bx')=(\kk^4\Ben\haut /\Per)f(\kk\haut) \;u_2(0,x_2)$ and $\kk^4\Ben=\rt\omega^2\cphi\Per$, we recover the form announced in \eqref{CT2}.
 \begin{figure}[h!]
\centering
\includegraphics[width=.5\columnwidth]{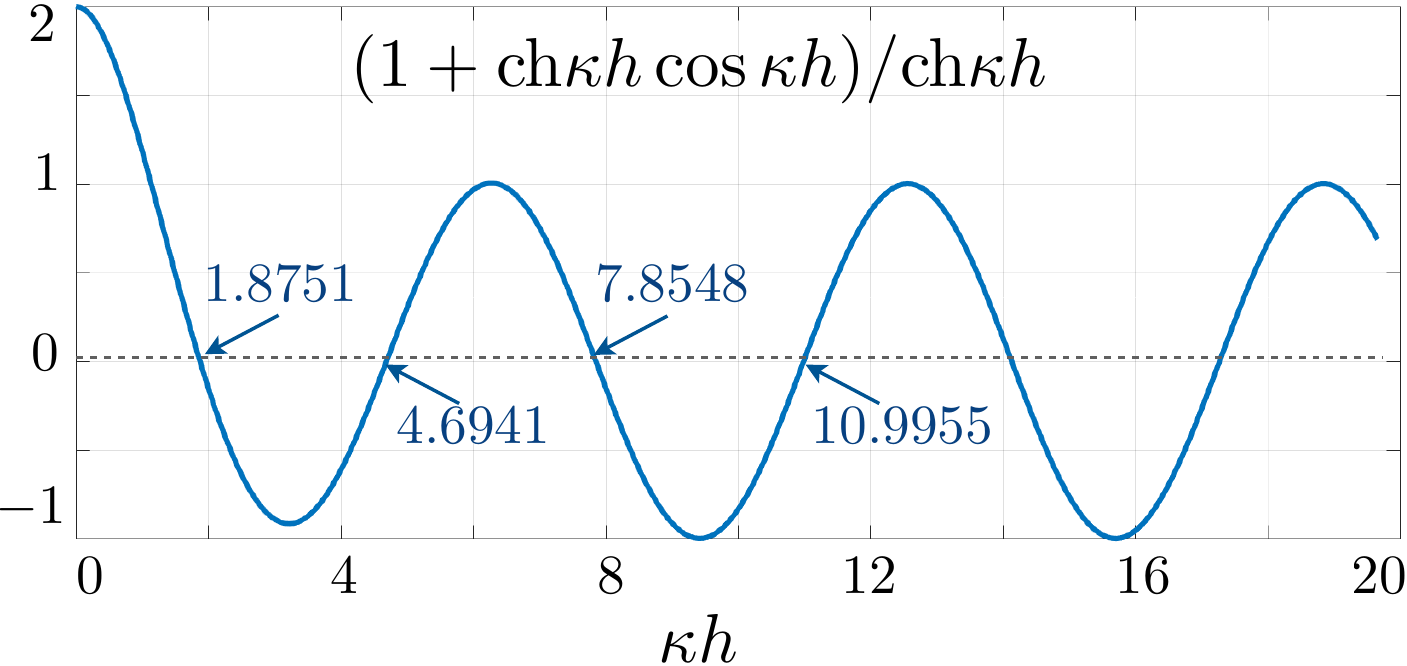}
	\caption{Bending resonances in the region of the plates (clamped/stress free conditions). }
	\label{Figa1}
\end{figure}
%


\section{Main steps of the derivation for an array of circular beams}
\label{appM}

Let us derive in this Appendix the effective model for circular beams of radius $\rp$, for which all the derivations are analytical; the circular beams are periodically located in a square array whose unit cell has a section area $S$ (figure \ref{Fig1}(b)). In this case, the complete formulation of the problem reads 
\beq\label{complet3}
\toutind
\dsp \text{In the substrate}, x_1\in(-\infty,0): & \div \sig+\rg \omega^2 \bu=\bo,  \quad \sig=2\mug \ee+\lamg \text{tr}(\ee) \mI,  \\[12pt]
\dsp \text{In the region of the beams}, x_1\in(0,\haut): &
\dsp \frac{\partial^4 u_{\alpha}}{\partial x_1^4}-\kk^4 u_\alpha=0,\quad \alpha=2,3,\quad
 \kk=\left( \frac{\rt \omega^2\pi\rp^2 }{\Ben} \right)^{1/4},
\toutout
\eeq
where  
\beq
 \Ben=\Et\,\frac{\pi \rp^4}{4},
 \eeq 
 is the flexural rigidity of the circular beams, complemented by
  the boundary conditions
\beq\label{CT3d}
\toutin
\dsp \sigma_{11}(0^-,\bx')=\rt\omega^2\cphi \haut \; u_1(0,x_2),\\[10pt]
\dsp \sigma_{1\alpha}(0^-,\bx')=-\frac{\Ben}{S}\,  \;\drt{u_\alpha}{x_1}(0^+,x_2), 
\\[10pt]
\dsp {u_\alpha}(0^+,\bx')={u_\alpha}(0^-,\bx'),\quad \dsp \dr{u_\alpha}{x_1}(0^+,\bx')=0,
\\[10pt]
\dsp \drd{u_\alpha}{x_1}(\haut,\bx')=\drt{u_\alpha}{x_1}(\haut,\bx')=0, \quad \alpha=2,3.
\toutout\eeq
where $\cphi=\pi \rp^2/S$. 
It follows that the   problem can be thought in the substrate only, with
\beq \label{gg3d}
\dsp \div \sig+\rg \omega^2 \bu=\bo,  \quad \sig=2\mug \ee+\lamg \text{tr}(\ee)\, \mI,  \quad x_1\in(-\infty,0),
\eeq
along with the boundary conditions of the Robin's type
\beq\label{CT3p}
\toutin
\dsp \sigma_{11}(0,\bx')=\; \epun u_1(0,\bx'),	  \\[10pt]
\dsp \sigma_{1\alpha}(0,\bx')
=\epun\;f(\kk\haut) \; u_\alpha(0,\bx'),
\toutout\eeq
where $Z$ and $f(\kk\haut)$ are defined in \eqref{deff} (we used that $\Ben \kk^4=\rt \omega^2\pi\rp^2$).

\subsection{Effective wave equation  in the region of the beams}
\subsubsection{Notations}

We shall use the same expansions as in \eqref{exp} but now, the terms $\bua^n$ and $\bsa^n$ depend on $\bz'=(z_2,z_3)$ (and not only on $z_2$) and we seek to establish the effective behaviour in the region of the array in terms of macroscopic averaged fields 
\beq
\mo{\bua^n}(y_1,\bx')=\frac{1}{\cphi \Ss}\int_{\Yt}\bua^n(y_1,\bz',\bx')\,\dd \bz', \quad \mo{\bsa^n}(y_1,\bx')=\frac{1}{\Ss}\int_Y\bsa^n(y_1,\bz',\bx')\,\dd \bz',
\eeq
where $\bx'=(x_2,x_3)$ and  $\Yt$ represents the circular section of the beam $\Yt=\left\{\sqrt{z_2^2+z_3^2}\leq \ray \right\}$, with $\ray=\rp/\ep^2$. It is worth noting that it is sufficient to replace $z_2$ by $\bz'$ in   \eqref{diff} to \eqref{dt}; in particular, we have 
\beq\label{diff2}
\grad\to \frac{\eu}{\ep}\dr{}{y_1}+ \frac{1}{\ep^2}\grad_{\bz'}+\grad_{\bx'}.
\eeq
and
\beq
 \ee^{\bz'}(\bua)=\frac{1}{2}\begin{pmatrix} 
0 & \frac{1}{2}\drs{\ua_1}{z_2} & \frac{1}{2}\drs{\ua_1}{z_3} \\[8pt]
\frac{1}{2}\drs{\ua_1}{z_2} & \drs{\ua_2}{z_2} & \frac{1}{2}\left(\drs{\ua_2}{z_3}+\drs{\ua_3}{z_2}\right) \\[8pt]
 \frac{1}{2}\drs{\ua_1}{z_3} & \frac{1}{2}\left(\drs{\ua_2}{z_3}+\drs{\ua_3}{z_2}\right) & \drs{\ua_3}{z_3}  
\end{pmatrix}.
 \eeq

\subsubsection{Sequence of resolution and main results of the analysis}
The analysis is made more involved since the problem is two-dimensional in the rescaled coordinate $\bz'$. 
The procedure is thus more complex.
It is as follow:

\begin{enumerate}
\item{ We establish that  
\beq\label{kim11}
\mo{\sa^0_{11}}=0, 
\eeq 
and  the dependence of $\bua^0$  on $(z_2,z_3)$, specifically
\beq\label{kim22}
\ua^0_1=W^0_1(\bx'), \quad \ua^0_\alpha=W^0_\alpha(y_1,\bx'), 
\eeq
}
\item{ We deduce the form of $\bsa^0$ and of $\bua^1$   
\beq\label{kim33}
\toutin
\dsp  \sa_{11}^0=- \Et \drd{W^0_\alpha}{y_1}(y_1,\bx')\; z_\alpha,\quad
\dsp   \sa_{1\alpha}^0=\sa_{\alpha\beta}^0=0,\\
\dsp \ua^1_1=W_1^1(\bx') - \dr{W^0_\alpha}{y_1}(y_1,\bx')\,z_\alpha,\quad \ua^1_\alpha=W^1_\alpha(y_1,\bx').
\toutout\eeq
}
\item{eventually the form of $\mo{\sa^1_{1i}}$ 
and the Euler-Bernoulli  equation for the bending  $W^0_\alpha$, $\alpha=2,3$. Specifically
\beq\label{kim44}\toutin
\dsp \mo{\sa_{11}^1}(y_1,\bx')=\rt \omega^2\cphi\, W_1^0(\bx')\,(\haute-y_1),\\[10pt]
\dsp \mo{\sa^1_{1\alpha}}(y_1,\bx')=- \Et\,\frac{\cphi\ray^2}{4}\drt{W_\alpha^0}{y_1}(y_1,\bx'),\toutout\eeq
and 
\beq\label{kim55}
\Et\,\frac{\ray^2}{4}\, \drq{\Ua^0_\alpha}{y_1} -\rt \omega^2\,\Ua^0_\alpha=0.
\eeq}
\end{enumerate}

\subsubsection{First step: $\mo{\sa^0_{11}}$ in \eqref{kim11} and  $\bua^0$ in \eqref{kim22}}
This step is not very demanding. From  $(\E_1)^{-1}$ in \eqref{eq13d},   
$\drs{\sa^0_{11}}{y_1}+\drs{\sa^1_{1\alpha}}{z_\alpha}=0$,
which after integration over $\Yt$ leaves us with $\partial_{y_1}\mo{\sa^0_{11}}=0$; anticipating  $\mo{\sa_{11}^0}=0$ at the top of the beams (as we did for the plates), we get  $\mo{\sa^0_{11}}=0$ everywhere, as announced in   \eqref{kim11}.

\vspace{.3cm}
Now, from  $(\C'_{1\alpha})^{-2}$ in \eqref{trez}, $\drs{\ua^0_1}{z_\alpha}=0$ and from $(\C'_{11})^{-1}$ 
$\partial_{y_1}\ua^0_1=0$. It follows that  $\ua_1^0$ depends only on $\bx'$, in agreement with \eqref{kim22}.
From $(\C'_{\alpha\beta})^{-2}$,  $\ua^0_\alpha$ is a rigid body motion \emph{i.e.} 
\beq\label{h1}
\ua^0_\alpha=W^0_\alpha(y_1,\bx')+\om^0(y_1,\bx')(\eu,\bz,\eal), 
\eeq
 with $(\eu,\bz,\eal)=\eu\cdot(\bz\times\eal)$ being the triple product, and we shall establish that $\om^0=0$. To do so we infer, from $(\C'_{1\alpha})^{-1}$, that 
\beq\label{h2}
 \partial_{z_\alpha}\ua^1_1+\partial_{y_1}\ua^0_\alpha=0  \rightarrow \partial_{y_1}\left(\partial_{z_2}\ua^0_3-\partial_{z_3}\ua^0_2\right)=0.
 \eeq
Inserting \eqref{h1} in \eqref{h2} tells us that 
 $\om^0$ does not depend on $y_1$ and anticipating the matching condition with the displacement in the substrate which imposes that $\om^0=0$ at $y_1=0$, we deduce that $\om^0=0$ everywhere, and \eqref{h1} reduces to the form of $\ua^0_\alpha$ announced in \eqref{kim22}.

\subsubsection{Second step:   ($\sa^0,\bua^1$) in \eqref{kim33}}
We start by determining $\bua^1$ incompletely (compared to what is announced in \eqref{kim22}). For $\ua_1^1$, we come  back to  $\partial_{z_\alpha}\ua^1_1+\partial_{y_1}\ua^0_\alpha=0$ in \eqref{h2}, and  $\ua^0_\alpha$ in \eqref{kim22} provides us with
\beq\label{moisi}
\ua_1^1=W_1(y_1,\bx')-\dr{W_\alpha^0}{y_1}(y_1,\bx')\,z_\alpha,
\eeq
and it remains for us to show that $W_1$ does not depend on $y_1$; this will be done after $\sa_{11}^0$ has been determined. 
Next, from $(\C'_{\alpha\beta})^{-1}$,  $\ua^1_\alpha$  is a rigid body motion, hence
\beq\label{moisi2}
\ua^1_\alpha=W^1_\alpha(y_1,\bx')+\om^1(y_1,\bx')(\eu,\bz,\eal).
\eeq
Now, we shall prove  that $\om^1=0$;  this will be done once $\sa^0_{1\alpha}$ have been determined. 
For the time being, we pursue  the calculations by  setting the boundary value problem   set in $Y$ on the unknowns $(\sa^0_{\alpha\beta},\ua^2_\alpha)$. From
$(\E_{\alpha})^{-2}$  and $(\C_{\alpha\beta})^0$ in \eqref{eq13d}, it reads 
\beq\toutin
 \partial_{z_\beta}{\sa^0_{\alpha\beta}}=0,\quad
 \sa^0_{\alpha\beta}=2\mut\left(\ee^{\bx'}_{\alpha\beta}(\bua^0)+\ee^{\bz'}_{\alpha\beta}(\bua^2)\right)+\lamt\left( \drs{\ua^1_{1}}{y_1}+\ee^{\bx'}_{\gamma\gamma}(\bua^0)+\ee^{\bz'}_{\gamma\gamma}(\bua^2) \right)\delta_{\alpha\beta},\quad \text{in } Y,\\[10pt]
  \sa^0_{\alpha\beta}n_\beta=0\quad \text{on}\quad \partial Y,\toutout
  \label{sne4}
\eeq
with $\bua^0$  known from \eqref{kim22} and $\ua^1_1$ from \eqref{moisi} at this stage. 
It is easy to check  that the solution of this boundary value problem is 
\beq\toutin
\sa^0_{\alpha\beta}=0,\\
\dsp \ua^2_\alpha=-e^{\bx'}_{\alpha\beta}(\bua^0)z_\beta-\frac{\lamt}{2(\mut+\lamt)}z_\alpha\dr{W_1}{y_1}(y_1,\bx')+\frac{\lamt}{2(\mut+\lamt)}g_\alpha+W^2_\alpha(y_1,\bx')+\om^2(y_1,\bx')(\eu,\bz,\eal),
\toutout
\label{dne4}
\eeq
where 
\beq
g_2= \left(\frac{z_2^2}{2}-\frac{z_3^2}{2}\right)\drd{W^0_{2}}{y_1}+z_2z_3\drd{W^0_{3}}{y_1}, \quad g_3= \left(\frac{z_3^2}{2}-\frac{z_2^2}{2}\right)\drd{W^0_{3}}{y_1}+z_2z_3 \drd{W^0_{2}}{y_1}.
\eeq
The above form of $\ua_\alpha^2$  along with $\ua_\alpha^0$ in \eqref{kim22} can now be  used to find $\sa^0_{11}=\left(2\mut +\lamt\right) \drs{w^1_{1}}{y_1}+\lamt\left(\drs{\ua^0_\alpha}{x_\alpha}+\drs{\ua^2_\alpha}{z_\alpha} \right)$, and we get
\beq\label{uyq5}
\sa^0_{11}=\Et\left(\dr{W_1}{y_1}(y_1,\bx')- \drd{W^0_{\alpha}}{y_1}(y_1,\bx')\,z_\alpha\right),
\eeq
where we used that $\Et=\mut(2\mut+3\lamt)/(\mut+\lamt)$. It is now sufficient to remember that $\mo{\sa^0_{11}}=0$ to get that $\drs{W_1}{y_1}=0$, hence the above expression of $\sa^0_{11}$ simplifies to the form announced in \eqref{kim33} and $\ua^1_1$ in \eqref{moisi} simplifies to to that in \eqref{kim33}.

\vspace{.4cm}

We now use the boundary value problem   set in $Y$ on the unknowns  $(\sa^0_{1\alpha},\ua^2_1)$.
From $(\E_{1})^{-2}$ and $(\C_{1\alpha})^{0}$ in \eqref{eq13d}, it reads as  follows
\beq\toutin
\drs{\sa^0_{1\alpha}}{z_\alpha}=0,\quad 
\sa^0_{1\alpha}=\mut\left(\drs{\ua^1_{\alpha}}{y_1}+\drs{\ua^0_{1}}{x_\alpha} +\drs{\ua^2_{1}}{z_\alpha} \right),\quad \text{in } Y,\\
  \sa^0_{1\alpha}n_\alpha=0,\quad \text{on}\quad \partial Y,
\toutout\label{aqa}\eeq
with $\ua^0_1$ known from \eqref{kim22}  and $\ua^1_\alpha$ from \eqref{moisi2} at this stage.
The solution is again found to be of the form
\beq \toutin
\dsp \sa^0_{1\alpha}=\mut \,\dr{\om^1}{y_1}(y_1,\bx')(\eu,\bz,\eal),\\
 \dsp \ua^2_1=W^2_1(y_1,\bx')-z_\alpha\left(\dr{W^1_{\alpha}}{y_1}(y_1,\bx')+\dr{W^0_{1}}{x_\alpha}(\bx')\right),\label{nht7}
\toutout\eeq
and we see that $\om^1=0$ implies $\sa^0_{1\alpha}=0$. 
To show that $\om^1=0$, we use   $
\drs{\sa^0_{1\alpha}}{y_1}+\drs{\sa^1_{\alpha\beta}}{z_\beta}=0$ which we infer from $(\E_{\alpha})^{-1}$. Multiplying by $\bv=v_\alpha \eal$ with the triple product $v_\alpha=(\eu,\bz,\eal)$ and integrating over $\Yt$, we find that
\beq
\int_{\Yt} v_\alpha\drs{\sa^0_{1\alpha}}{y_1} \; \dd \bz'+\int_{\partial \Yt} v_{\alpha}\sa^1_{\alpha\beta}n_\beta  \;\dd l -\int_{\Yt}  \drs{v_{\alpha}}{z_\beta} \sa^1_{\alpha\beta}\;\dd \bz'=0.\label{kzo9}
\eeq
Since $\bsa^1$ is symmetric and $\grad \bv$ is anti-symmetric, we have $ \drs{v_{\alpha}}{z_\beta} \sa^1_{\alpha\beta}=0$, and  $\sa^1_{\alpha\beta}n_\beta=0$ on  $\partial \Yt$. Hence, \eqref{kzo9} reduces to
\beq
\int_{\Yt} v_\alpha\dr{\sa^0_{1\alpha}}{y_1} \; \dd \bz'=0 \rightarrow \drd{\om^1}{y_1}(y_1,\bx') \int_Y (\eu,\bz,\eal)^2 \dd\bz'=0.
\label{kzo10}
\eeq
Next, with $(\eu,\bz,\eal)^2=\bz^2$ whose integral does not vanish, we obtain that $\drs{\om^1}{y_1}$ does not depend on $y_1$;  anticipating that $\drs{\om^1}{y_1}(\haute,\bx')=0$  and   $\om^1(\haute,\bx')=0$, we deduce that $\om^1=0$ everywhere. 
It follows that $\sa^0_{1\alpha}=0$, from \eqref{nht7}, and that $\ua^1_\alpha=W^1_\alpha(y_1,\bx')$, from \eqref{moisi2}, in agreement with \eqref{kim33}.

\subsubsection{Third step: $\mo{\sa_{1i}^1}$ in \eqref{kim44} and  the Euler-Bernoulli equations in \eqref{kim55}}
This  starts with   $(\E)^0$ in \eqref{eq13d} integrated over $Y$, specifically
\beq
\dsp \dr{\mo{\sa^1_{11}}}{y_1}+\rt\omega^2\cphi\,W_1^0=0,\quad
\dsp \dr{\mo{\sa^1_{1\alpha}}}{y_1}+\rt\omega^2\cphi\,W_\alpha^0=0,
\label{yyy3}\eeq
where we have used that $\mo{\bsa^0}=0$ from \eqref{kim11} and \eqref{kim33} and $\bsa^2{\bf n}_{|\bY}=\bo$. Since $W_1^0$ depends only on $\bx'$, and
anticipating that $\mo{\sa^1_{11}}(\haute,\bx')=0$, we obtain by integration the form of $\mo{\sa^1_{11}}$ in \eqref{kim44}. 

\vspace{.3cm}

To get $\mo{\sa_{1\alpha}^1}$, we  multiply $(\E_1)^{-1}$  (which reads  $\drs{\sa^0_{11}}{y_1}+\drs{\sa^1_{1\beta}}{z_\beta}=0$, with $\sa^0_{11}$ in \eqref{kim33})  by $z_\alpha$ and integrate over $\Yt$ to find that 
\beq
 \Et\drt{W^0_{\beta}}{y_1}(y_1,\bx')\int_{\Yt} z_\alpha z_\beta \; \dd s+\int_{\Yt}\sa^1_{1\beta}\,\delta_{\alpha\beta} \; \dd s=0,
\eeq
where we have used that $ {\sa^1_{1\beta}\,n_\beta}_{|\bY} =0$. For the circular cross-section of the beams, $\int_Y z_2z_3\dd \bz'=0$ and $\int_Y z_2^2\dd \bz'=\int_Y z_3^2\dd \bz'=\pi\ray^4/4$. It follows that 
\beq
\Ss \mo{\sa^1_{1\alpha}}(y_1,\bx')=-\Et\frac{\pi\ray^4}{4}\,\frac{\partial^3 W^0_{\alpha}}{\partial y_1^3}(y_1,\bx'),\label{kzc5}
\eeq
in agreement with \eqref{kim44} (with $\cphi \Ss=\pi\ray^2$). Coming back to \eqref{yyy3} with the above form of $\mo{\sa^1_{1\alpha}}$, we  deduce that
\beq
\Et\frac{\pi\ray^4}{4}\,\frac{\partial^4 W^0_{\alpha}}{\partial y_1^4}-\rt\omega^2 \cphi\Ss W^0_\alpha=0,\label{kot3}
\eeq
in agreement with \eqref{kim55}.

\subsection{ Effective boundary conditions at the top of the array of beams} 

As we have done in \eqref{toto1}, we consider the following expansions  for the displacement and stress
\beq\label{lam1}
\bu=\sum_{n\geq0}\ep^n\bv^n(\bz,\bx'), \quad \sig=\sum_{n\geq 0}\ep^n\btau^n(\bz,\bx'). 
\eeq
We use $(\e)$ in  \eqref{eq11} (with $\bz'\to \bz$) which  provide us with
$\div_\bz \btau^0=\div_\bz \btau^1=\bo$, and this makes the calculations identical to those conducted in \S\ref{eff2} for the plates when integrating over $\Zt=\left\{z_1\in(-\infty,0), \bz'\in \Yt\right\}$. We thus obtain 
\beq
\mo{\sa^0_{1i}}(\haute,\bz',\bx')= \mo{\sa^1_{1i}}(\haute,\bz',\bx')=0, \quad i=1,2,3,
\eeq
(see \eqref{gg}).
The  conditions on $\mo{\sa^0_{1i}}$ are consistent with \eqref{kim11} and \eqref{kim33}. The condition on  $\mo{\sa_{11}^1}$ is that anticipated  to find \eqref{kim44}. Eventually, the condition on $\mo{\sa_{1\alpha}^1}$  combined with \eqref{kim44}  provides the conditions of zero shear force  
\beq
\drt{\Ua_\alpha^0}{y_1}(\haute,\bx')=0.\label{nbz0}
\eeq
To derive the condition of zero bending moment, we proceed the same as we have done in \eqref{kzo9}; with $\bv=v_\alpha \eal$ and  $v_\alpha=(\eu,\bz,\eal)$, we consider the vanishing  integral $\int_{\Zt} v_\alpha \partial_{z_i} \tau_{i\alpha}^0\; \dd v=0$, hence
\beq
\int_{\partial Z} v_\alpha \tau^0_{i\alpha}n_i=0,
\eeq
where we have used that $\partial_{z_i}v_\alpha \tau^0_{i\alpha}=0$ by construction. Because   $\btau^0 \cdot{\bf n}$ vanishes on $\partial \Zt$ except at the bottom face $z_1=-z_m$ and passing to the limit $z_m\to +\infty$, this integral reduces to 
\beq
\int_{\Yt} v_\alpha\, \sa^0_{1\alpha}(\haute,\bz',\bx') \; \dd s=0\label{neh7}.
\eeq
Making use of  \eqref{nht7}  leads to  the anticipated boundary condition
\beq
\dr{\om^1}{y_1}(\haute,\bx')=0,
\eeq
that we have used to get $\sa_{1\alpha}^0=0$.
 It remains to derive the  condition of zero bending moment. By considering $\ba=z_\alpha\,\eu-z_1\eal$ and integrating over $Z$  the scalar   $\ba\cdot \div_\bz \btau^0$ (since $\div_\bz \btau^0=\bo$), we found that
\beq
0=\int_{\partial Z}a_i \;\tau^0_{ij} n_{j} \;\dd s=-\int_Y \,z_\alpha\,{\tau^0_{11}}_{|{z_1=-z_m}} \;\dd s-z_m\int_Y \,{\tau^0_{1\alpha}}_{|{z_1=-z_m}} \;\dd s.
\eeq
Since we have in addition $0=\int_{\partial Z}\,\tau^0_{ij} n_{j} \;\dd s=\int_Y \,{\tau^0_{1\alpha}}_{|{z_1=-z_m}} \;\dd s$, we can pass to the limit $z_m\to \infty$, and get $0=\int_Y z_\alpha \tau^0_{11}\to \mo{z_\alpha\sa^0_{11}}(\haute,\bx')$. Now accounting for $\sa^0_{11}$ in  \eqref{uyq5}, we obtain the expected boundary condition
\beq
\drd{\Ua_\alpha^0}{y_1}(\haute,\bx')=0.\label{aal1}
\eeq

\subsection{Effective transmission conditions  between the substrate and the  array} 
In the vicinity of  the interface between  the substrate and the array, we consider the same expansions as in \eqref{lam1}, and at the dominant order, we still have 
$\div_\bz \btau^0=\div_\bz \btau^1=\bo$. The calculations are identical to that conducted in \S \ref{eff3} when integrating over $Z=\{z_1\in(0,+\infty), \bz'\in \Yt\}\cup\{z_1\in(-\infty,0), \bz'\in(-\per/2,\per/2)^2\}$, and we find 
\beq\label{ghf2}
 \sigma_{1i}^0(0^-,\bx')=0, \quad i=1,2,3,
 \eeq
 which are consistent with \eqref{kim11} and \eqref{kim33}. 
 Next, using $\mo{\sa^1_{1i}}$ in \eqref{kim44}, we find 
\beq\label{ghf1}
\sigma^1_{11}(0^-,\bx')=\rt\omega^2\,\cphi\, \haute W^0_1(\bx'),\quad \sigma^1_{1\alpha}(0^-,\bx')=-\Et \frac{\pi\ray^4}{4S}\frac{\partial^3 W^0_{\alpha}}{\partial y_1^3}(0,\bx').
\eeq
We have yet to establish the continuity of the displacement. From the counterpart of $(\cc')$ in \eqref{kim} (with $\bz'\to {\bz}$), it is easily seen that we have at the dominant orders 
\beq
\ee^\bz(\bv^0)=\ee^\bz(\bv^1)=\bo.
\eeq
Therefore $\bv^0$ and $\bv^1$ are piecewise rigid body motions, namely $\bv^0={\bm \Omega}^0(\bx')\times \bz +{\bf V}^0(\bx')$, the same for $\bv^1$. Invoking the periodicity of $\bv^i$, $i=1,2$ with respect to $z_2$ and $z_3$ for $z_1<0$ and the continuity of $\bv^i$ at $z_1=0$, these rigid body motions reduce to a single translation over $Z$. Eventually, using the matching conditions on the displacements, we then obtain
\beq\label{gfh}
u^0_1(0^-,\bx')=W^0_1(\bx'),\quad u^0_\alpha(0^-,\bx')=W^0_\alpha(0^+,\bx'),\quad \dr{W^0_\alpha}{z_1}(0^+,\bx')=0.
\eeq

\subsection{The final problem}
The effective problem \eqref{complet3} is obtained for $(\bu=\bu^0,\sig=\sig^0+\ep\sig^1)$  in the substrate for $x_1<0$, $(\bu={\bf W}^0,\sig=\bsa^0+\ep\bsa^1)$ in the region of the array for $x_1>0$. Remembering that   $y_1=x_1/\ep$ and $\haute=\haut/\ep$, $\ray=\rp/\ep^2$, it is easy to see that (i) the Euler-Bernoulli equation in \eqref{complet3} is obtained from \eqref{kim55}, (ii) the effective boundary conditions announced in \eqref{CT3d} from \eqref{nbz0}, \eqref{aal1}-\eqref{ghf1} and   \eqref{gfh}.

\section*{References}

\bibliography{Nmybibfile}

\end{document}